\RequirePackage{luatex85}
\documentclass[%
 amsmath,amssymb,
 aps,
]{revtex4-2}

\usepackage{graphicx}
\usepackage{dcolumn}
\usepackage{bm}

\usepackage{xcolor}
\usepackage{graphicx}
\usepackage{float}
\usepackage{epstopdf, epsfig}
\usepackage{tikz}
\usepackage[utf8]{inputenc}
\usepackage{pgfplots}

\pgfplotsset{plot coordinates/math parser=false}
\usetikzlibrary{arrows.meta}
\pgfplotsset{compat=1.10}
\usepgfplotslibrary{fillbetween}
\usepackage{epsfig}
\usepackage[english]{babel}
\pgfplotsset{compat=newest}
\usetikzlibrary{external}
\tikzsetexternalprefix{tikzExternalized/}
\tikzset{external/force remake}
\usepackage{shellesc}

\begin{document}

\preprint{APS/123-QED}

\title{Passive stabilization of crossflow instabilities by a reverse lift-up effect}

\author{Jordi Casacuberta}
\author{Stefan Hickel}
\author{Marios Kotsonis}
\affiliation{Delft University of Technology, Delft 2629HS, The Netherlands}%

\date{October 4, 2023}

\begin{abstract}
A novel mechanism is identified, through which a spanwise-invariant surface feature (a two-dimensional forward-facing step) significantly stabilizes the stationary crossflow instability of a three-dimensional boundary layer. The mechanism is termed here as \textit{reverse lift-up effect}, inasmuch as it acts reversely to the classic lift-up effect; that is, kinetic energy of an already existing shear-flow instability is transferred to the underlying laminar flow through the action of cross-stream perturbations. To characterize corresponding energy-transfer mechanisms, a theoretical framework is presented, which is applicable to generic three-dimensional flows and surface features of arbitrary shape with one invariant spatial direction. The identification of a passive geometry-induced effect responsible for dampening stationary crossflow vortices is a promising finding for Laminar Flow Control applications. 
\end{abstract}


\maketitle

\section{Introduction}

The understanding of interactions between boundary-layer instabilities and surface features is pivotal for laminar-turbulent transition research. For decades, the aerodynamics community has devoted many efforts to the characterization of mechanisms by which distributed (i.e.\ invariant in one spatial direction), three-dimensional isolated or in array configuration, and other kinds of surface features alter the transition route. With very few exceptions \cite{Fransson06}, general consensus dictates that rapid spatial variations of the surface geometry advance the transition front upstream. In this regard, this work identifies a novel flow mechanism by which a local modification of the surface geometry (in this case a distributed forward-facing step) actually stabilizes a convective instability in subsonic swept-wing flow. This finding complements recent experimental investigations in the authors' group at TU Delft which demonstrated empirically that forward-facing steps have the capability to delay the laminar-turbulent transition of boundary \mbox{layers \cite{Rius21}}. To delay transition by passive means (i.e.\ no energy input required) is a major ambition in the field of Laminar Flow Control and motivated by the significantly lower skin-friction drag of a laminar boundary layer, as opposed to a turbulent one. Overall, the framework presented in this work aims at setting theoretical foundations for further research on passive laminarization of swept aerodynamic bodies by suitable design of surface relief.

The published literature on the effects of spanwise-invariant surface features on transition elaborates mainly on the manner by which transition is actually promoted. From early experiments on two-dimensional (i.e.\ unswept) flow, it was established that the transition-front location is bounded between the surface feature and the transition-front location in reference (i.e.\ no-surface-feature present) conditions \cite{Fage41,Tani69,Dryden53}. Much of the available empirical knowledge at the time materialized essentially in a so-called roughness-equivalent Reynolds number, often denoted as $Re_{k}$ or $Re_{h}$, which became a popular transition-correlation parameter \cite{Smith59,Braslow60}. Concerning the main flow mechanism(s) by which a surface feature promotes transition, two essential agents have been debated historically. On the one hand, the destabilizing influence of the mean-flow profiles in the wake of the surface feature; see for instance the discussion by Klebanoff and Tidstrom \cite{Klebanoff72}. On the other hand, perturbations introduced by the surface feature itself; in this regard, the work of Goldstein \cite{Goldstein85} is highlighted, who investigated the coupling between an incoming Tollmien-Schlichting (TS) instability and acoustic disturbances scattered at the surface feature \cite{Saric02}. Notwithstanding this predominant role of spanwise-invariant surface features in promoting transition, there are reports of successful boundary-layer stabilization by smooth protuberances in hypersonic-flow applications \cite{Fujii06,Duan13,Fong14,Park16}.

The scope of this work is specifically placed on subsonic three-dimensional swept-wing flow and a transition route initiated by the amplification of a (stationary) crossflow eigenmode. In subsonic conditions, swept-wing flow is susceptible to four main instability kinds: attachment line, Görtler, TS, and crossflow \citep{Saric03}. Large crossflow instability (CFI) growth is expected in regions of strong favorable pressure gradient, as for instance, near the leading edge \cite{Arnal93}. In this flow regime, the existence of an inflectional boundary-layer profile in the direction approximately orthogonal to the trajectory of inviscid streamlines sets the conditions for modal (crossflow) instability growth. While this instability mechanism may manifest either as a stationary (i.e.\ of zero temporal frequency) or a traveling (i.e.\ of non-zero temporal frequency) wave-like perturbation, its stationary form dominates in free-flight conditions characteristic of low-turbulence environments \cite{Saric03}. Oppositely, traveling crossflow is favored typically in high-turbulence environments in combination with high levels of surface finish, that is, smooth surfaces \cite{Deyhle96}. When superimposed on the time-invariant laminar swept-wing base flow, the spatial growth of the stationary CFI reveals as co-rotating (crossflow) vortices that are distributed periodically along the leading-edge-parallel (or, spanwise) direction and increase in strength when moving along the leading-edge-orthogonal (or, chordwise) direction. However, at perturbation level, i.e.\ when isolated from the underlying base flow, the stationary CFI develops as a wave-like cross-stream pattern that, by displacing $\mathcal{O}(1)$ streamwise momentum, enhances regions of streamwise-momentum deficit and excess \cite{Saric03}. This rather weak cross-stream pattern appears structurally as vortical perturbation structures, typically referred to as rolls, that counter-rotate with respect to each other. Following stages of linear and subsequent non-linear perturbation amplification, the development of unsteady secondary instabilities on the shear layers embedding the stationary crossflow vortices leads ultimately to laminar breakdown \cite{Saric03,Deyhle96,Bippes99,Malik99,Haynes00,Wassermann02,Serpieri16}.

The need to understand how this \textit{classic} transition route is altered on realistic engineering surfaces containing irregularities has motivated a wealth of numerical and experimental studies in recent years. Particularly, much attention has been devoted to the mechanisms of interaction between a pre-existing or incoming stationary CFI and a spanwise-distributed forward-facing step \cite{Tufts17,Eppink20,Rius21,Rius22,Casacuberta22JFM}. The current consensus establishes that the step amplifies locally the incoming instability upon interaction, when compared to a reference no-step scenario. Tufts \textit{et al.} \cite{Tufts17} argue that, when a particular step height is exceeded, the stationary CFI grows sharply while passing over the step. They attribute the effect to a constructive \textit{interaction} between the incoming crossflow vortices and recirculating flow on the upper wall of the step. Eppink \cite{Eppink20} describes strong stationary CFI growth locally at the step, followed by a stage of decay and further growth when moving downstream of it; the destabilizing influence of step-induced inflectional profiles near the wall is deemed responsible mainly for the initially monitored phase of stationary CFI amplification. Rius-Vidales and Kotsonis \cite{Rius21,Rius22} study experimentally the influence of the step's height on the transition-front location. Laminar-turbulent transition is found to be almost universally promoted for most considered step height and incoming perturbation amplitude combinations. For the smallest step studied, however, an unexpected transition delay is found \cite{Rius21}. Comparably, it has been shown that shallow surface strips, i.e.\ forward- and backward-facing steps, may delay crossflow-induced transition by reducing the strength of crossflow vortices \cite{Ivanov19}. 

Casacuberta \textit{et al.} \cite{Casacuberta22JFM} perform steady-state Direct Numerical Simulations (DNS) and observe that the pre-existing stationary CFI of primary wavelength, i.e.\ the integrally most amplified mode, can be significantly stabilized locally downstream of the step for particular choices of the step height. This apparent discrepancy with the main consensus in that forward-facing steps may not always destabilize stationary crossflow vortices is ascribed to the fact that additional velocity-perturbation streaks that are induced at the upper step corner as a by-product of the interaction \cite{Casacuberta21,Casacuberta22JFM}. The sharp growth of these locally-formed perturbation streaks obscures the fact that the incoming instability --which develops above the streaks while passing over the step-- undergoes non-modal growth and is actually stabilized for particular choices of step height \cite{Casacuberta22JFM}. It has been described above that the amplification of crossflow vortices in reference (i.e.\ no-step) conditions is driven essentially by a modal perturbation effect. However, three-dimensional swept-wing flow is prone to significant non-modal growth as well \cite{Breuer94}. More specifically, it has been shown that the crossflow-perturbation wave is the wave type which yields the largest non-modal growth \cite{Breuer94}; thus, modal- and non-modal-growth mechanisms exhibit a structurally similar perturbation response in a swept-wing boundary layer \cite{Breuer94,Corbett01,Tempelmann10}. In this regard, it is noted by Breuer and Kuraishi \cite{Breuer94} that a non-modal-growth mechanism --triggered for instance by a localized surface feature-- results in the inception of streamwise streaks. In unswept forward-facing-step flows, velocity-perturbation streaks structurally similar to those reported by Casacuberta \textit{et al.}  \cite{Casacuberta21,Casacuberta22JFM} have been identified as well and ascribed to the lift-up effect \cite{Lanzerstorfer12}.  

The so-called lift-up effect \cite{Moffatt67,Ellingsen75,Landahl75,Landahl80} is considered a key mechanism behind the amplification of perturbations in shear flow since its formalization in the $1970$s and $1980$s. Previously, more than a century ago, Lord Rayleigh's Inflection Point Theorem \cite{Rayleigh80} established a main result of hydrodynamic stability. Namely, a necessary condition for instability (i.e.\ exponential growth of wave-like perturbations) of parallel incompressible inviscid two-dimensional flow is that the mean-velocity profile possesses an inflection point. However, wide experimental evidence has repeatedly shown that perturbation growth can occur also in absence of inflectional mean-flow profiles and that laminar-turbulent transition takes place in scenarios where linear modal stability analysis predicts stable flow. The investigations of Ellingsen and Palm \cite{Ellingsen75} and Landahl \cite{Landahl75,Landahl80} --on what later was referred to as the lift-up effect-- aided to address this apparent contradiction. Their work essentially establishes that, from the viewpoint of perturbation kinetic energy, any parallel inviscid shear flow is unstable to a large set of initial three-dimensional perturbations. This is irrespective of the exponential (or, asymptotic modal) stability of the flow. The physical principle of the lift-up effect is illustrated typically as follows: a cross-stream perturbation of wave-like form superimposed on a shear layer \textit{lifts up} low-momentum fluid and pushes down high-momentum fluid in adjacent regions of the flow field following the wave-like distribution. By retaining their original streamwise momentum, the displaced fluid particles introduce regions of streamwise-momentum deficit and excess and therefore induce inherently streamwise \textit{perturbation} streaks that can attain very large amplitude in a short span of space or time. By this principle, the lift-up effect is associated to the mechanism of \textit{optimal} streamwise vortices \cite{Andersson99,Luchini00} (i.e.\ the initial perturbation yielding the largest possible \textit{transient growth} of kinetic perturbation energy) relaxing into streamwise streaks following a non-modal (or, algebraic) growth \cite{Butler92,Reddy93,Schmid01,Schmid07}. The lift-up effect is also known to play a key role in bypass transition of boundary-layer flows subject to high levels of free-stream turbulence or surface roughness \cite{Klebanoff71,Kendall85,Westin94,Andersson01,Matsubara01,Jacobs01,Reshotko01,Brandt03,Brandt04,Zaki05,White05b,Brandt07}.   Overall, the lift-up effect is presently regarded as a major mechanism responsible for the ubiquitous presence of streaky structures in shear flow \cite{Brandt14}. 

The lift-up effect was originally formulated by considering inviscid two-dimensional parallel flow with a streamwise-independent perturbation \cite{Ellingsen75}. Throughout the years, more sophisticated frameworks interpret and quantify the underlying phenomenon in more generic flow environments; for instance, by considering a vorticity-perturbation formulation, see the illustrative example by Roy and Subramanian \cite{Roy14} in their figure $1$. Recently, it has been proposed to characterize the lift-up effect through the production term of the Reynolds-Orr equation; that is, the equation governing the evolution of kinetic perturbation energy \cite{Schmid01}. Particularly, following the methodology of Albensoeder \textit{et al.} \cite{Albensoeder01}, Lanzerstorfer and Kuhlmann \cite{Lanzerstorfer11,Lanzerstorfer12} express this production term into four main contributions, which result from the decomposition of the perturbation vector into components locally tangential and normal to base-flow streamlines. A main term arising from the decomposition of production, typically referred to as $I_{2}$, characterizes the lift-up effect \cite{Lanzerstorfer11,Lanzerstorfer12,Loiseau16,Picella18}. That is, in certain flow environments, $I_{2}$ expresses the transfer rate of kinetic energy between the underlying laminar base flow and the streamwise-velocity (i.e.\ flow aligned) perturbation by the action of the cross-stream-velocity (i.e.\ flow orthogonal) perturbation on the base-flow shears. 

While the approach of Albensoeder \textit{et al.} \cite{Albensoeder01} and Lanzerstorfer and Kuhlmann \cite{Lanzerstorfer11,Lanzerstorfer12} provides an attractive way to characterize the lift-up effect in front of other more convoluted methods, several related critical points demand further exploration; we highlight two major ones next. First, in their original studies \cite{Albensoeder01,Lanzerstorfer11,Lanzerstorfer12}, the reference laminar base flow on which perturbation evolution is analyzed is two-dimensional. Formulation and interpretation of the underlying ideas considering more generic three-dimensional flow is necessary to fully generalize the method and equations. Second, it is well known that the sign of the production term of the Reynolds-Orr equation establishes the \textit{sense} of kinetic energy exchange; that is, whether kinetic energy is transferred from the base flow to the perturbation field and the process acts destabilizing, or vice-versa. This interpretation is identically applicable to the individual terms arising from the decomposition of the production term \cite{Albensoeder01}. Therefore, it follows from this mathematical rationale that the lift-up effect, $I_{2}$, may as well act towards stabilizing the flow through an inversion of sign. This is in apparent contradiction to the common conception in the literature that the lift-up effect is a destabilizing flow mechanism responsible for the inception of highly energetic streamwise streaks.

This work is motivated by the identification of a \textit{reverse} (i.e.\ stabilizing) \textit{lift-up effect} and its key role played in novel passive stabilization of crossflow instability by a local modification of surface geometry (namely a spanwise-invariant forward-facing step) \cite{Casacuberta22JFM}. Specifically, the goal of this article is threefold: (i) to illustrate that the lift-up effect, characterized through the production term of the Reynolds-Orr equation, quantifies major perturbation effects in the interaction between pre-existing stationary crossflow instability and forward-facing steps of several heights. Pertinent numerical investigations are addressed by means of steady-state Direct Numerical Simulations. (ii) To illustrate and generalize that, under certain conditions, the lift-up effect may act stabilizing and thus lead to decay of kinetic perturbation energy for a finite spatial extent. During this process, the associated mathematical term reverses in sign and the underlying mechanism of the lift-up effect acts essentially reversely to the classic conception introduced by Ellingsen and Palm \cite{Ellingsen75} and Landahl \cite{Landahl75,Landahl80}. This manifests by quenching of pre-existing streamwise streaks through the displacement of low- and high-momentum fluid by a cross-stream perturbation. (iii) To bring forward that the mechanism by which a forward-facing step of particular height stabilizes locally a stationary crossflow instability \cite{Casacuberta22JFM} and leads potentially to passive delay of laminar-turbulent transition in swept-wing flow \cite{Rius21} is a reverse lift-up effect. The methodology presented in this article may be extended to other classes of shear-flow instabilities and surface features of arbitrary shape with one invariant spatial direction. In summary, the current work aspires to provide an analysis and design methodology for the understanding of destabilizing and stabilizing effects of surface features in critical transitional flows. This can be further exploited to design and optimize passive laminar flow control strategies.

The work is structured as follows. Section \ref{sec:methodology} provides analytical expressions describing the lift-up effect and the cross-stream component of a three-dimensional perturbation field in a spanwise-invariant base flow. This section describes additionally the main flow problem discussed in this article, i.e.\ crossflow instability interacting with step flow, together with two model problems employed towards illustrating the concept of a reverse lift-up effect. Namely, \textit{optimal} perturbations in plane Poiseuille flow and wall blowing-suction in a two-dimensional boundary-layer flow. These two model problems are analyzed in Sections \ref{sec:modelI} and \ref{sec:modelII}. Section \ref{sec:resultStep} presents the main results of the step-flow problem: first, it describes the perturbation evolution at the step. Then, it characterizes mathematically the lift-up effect at the step and the conditions under which it may act towards stabilising or destabilising the flow. Finally, a physical interpretation is given for the reversed action of the lift-up effect. Section \ref{sec:conclusions} states the conclusions. 

\section{Methodology}\label{sec:methodology}

\subsection{Projection of the perturbation field to the local base flow}\label{sec:mainFormulation}

A theoretical framework is introduced in this article to scrutinize perturbation evolution and the mechanisms of kinetic-energy exchange between perturbation fields and an underlying laminar (unperturbed) base flow. The framework is applicable to generic three-dimensional base flows with one invariant spatial direction. Considering the main flow problem of this work, a stationary (i.e.\ time-independent) perturbation field is assumed. However, extension to unsteady perturbations is straightforward.

A Cartesian coordinate system is considered, $\bm{x} = [x \; y \; z]^{\textrm{T}}$, that is fixed and oriented relative to a reference geometry of the flow problem; that is, $y$ expresses the wall-normal direction, $z$ is the spanwise (or, transverse) direction, and $x$ completes the coordinate system pointing typically in the main direction of the flow. A three-dimensional unperturbed laminar base flow is assumed, which is invariant in one spatial direction --the spanwise direction $z$ in this work-- and whose velocity vector reads $\bm{\upsilon}_{\textrm{B}} = [u_{\textrm{B}} \; v_{\textrm{B}} \; w_{\textrm{B}}]^{\textrm{T}}$ with $\partial \bm{v}_{\textrm{B}} / \partial z = 0$. For the derivation below, it is important that $z$ is the direction of statistically homogeneous flow. For the reminder of this article, $u$, $v$, and $w$ denote velocity components in the $x$, $y$, and $z$ directions, respectively. The superposition of a three-dimensional stationary velocity-perturbation field, $\hat{\bm{v}}^{\prime}$ with the unperturbed base flow forms a \textit{steady developed flow}; i.e.
\begin{equation}\label{eq:BFandPertDecomp}
\bm{v}(x,y,z) = \bm{v}_{\textrm{B}}(x,y) + \hat{\bm{v}}^{\prime}(x,y,z).
\end{equation}

\noindent The perturbation field $\hat{\bm{v}}^{\prime}= [\hat{u}^{\prime} \; \hat{v}^{\prime} \; \hat{w}^{\prime}]^{\textrm{T}}$ is conceived as spanwise-periodic and, as such, it is amenable to wave-like representation:
\begin{equation}\label{eq:pertAnsatz}
\hat{\bm{\upsilon}}^{\prime}(x,y,z) = \underbrace{\tilde{\bm{\upsilon}}(x,y) \ \textrm{e}^{ \textrm{i} \beta_{0} z}}_{\bm{\upsilon}^{\prime}} \; + \; \textrm{c.c.},
\end{equation}

\noindent where $\tilde{\bm{\upsilon}} \in \mathbb{C}^{3}$ expresses the Fourier (or, amplitude) coefficient, $\beta_{0}$ is the fundamental spanwise wavenumber, $\textrm{c.c.}$ stands for complex conjugate (also denoted by $\dagger$ in this article), and $\textrm{i}^2 = -1$. 

The modulus of the components of the Fourier coefficient $\tilde{\bm{\upsilon}}$ read $|\tilde{u}|$, $|\tilde{v}|$, and $|\tilde{w}|$ and are referred typically to as amplitude functions; the phase (or angle) of each component is denoted respectively by $\varphi^{u}$, $\varphi^{v}$, and $\varphi^{w}$. Following the approach by Albensoeder \textit{et al.} \cite{Albensoeder01} and Lanzerstorfer and Kuhlmann \cite{Lanzerstorfer11,Lanzerstorfer12}, we decompose $\bm{\upsilon}^{\prime}$ relative to the orientation of the base flow \citep{Albensoeder01,Marxen09,Lanzerstorfer11,Lanzerstorfer12,Picella18,Casacuberta22JFM} instead of relative to the orientation of the wall:
\begin{equation}\label{eq:pertDecompBF}
\bm{\upsilon}^{\prime}(x,y,z) = \bm{\upsilon}_{t}^{\prime}(x,y,z) + \bm{\upsilon}_{n}^{\prime}(x,y,z) = [\upsilon^{\prime 1}_{t} \; \upsilon^{\prime 2}_{t} \; \upsilon^{\prime 3}_{t}]^{\textrm{T}} + [\upsilon^{\prime 1}_{n} \; \upsilon^{\prime 2}_{n} \; \upsilon^{\prime 3}_{n}]^{\textrm{T}}.
\end{equation}

\noindent The field $\bm{\upsilon}_{t}^{\prime}$ in Eq.~(\ref{eq:pertDecompBF}) characterizes the content of $\bm{\upsilon}^{\prime}$ which acts tangential to the base flow. That is, the perturbation which, at every point in space, points in the local streamwise direction $\hat{\boldsymbol{t}} = \bm{\upsilon}_{\textrm{B}} / ||\bm{\upsilon}_{\textrm{B}}||$:
\begin{equation}\label{eq:tanDef}
\bm{\upsilon}_{t}^{\prime} = \tau^{\prime}\hat{\boldsymbol{t}}
\end{equation}

\noindent with
\begin{equation}\label{componentTan}
\tau^{\prime} = \frac{1}{||\bm{\upsilon}_{\textrm{B}}||} \sqrt{\left( \gamma^{+} \right)^2 + \left( \gamma^{-} \right)^2} \; \textrm{e}^{\textrm{i} \left( \beta_{0}z + \varphi_{t} \right)},
\end{equation}
\begin{equation}
\begin{split}
&\gamma^{+}(x,y) = u_{\rm{B}} |\tilde{u}| \cos(\varphi^{u}) + v_{\rm{B}} |\tilde{v}| \cos(\varphi^{v}) + w_{\rm{B}} |\tilde{w}| \cos(\varphi^{w}), \\
&\gamma^{-}(x,y) = -u_{\rm{B}} |\tilde{u}| \sin(\varphi^{u}) - v_{\rm{B}} |\tilde{v}| \sin(\varphi^{v}) - w_{\rm{B}} |\tilde{w}| \sin(\varphi^{w}),
\end{split}
\end{equation}

\noindent and
\begin{equation}\label{eq:phaseTang}
\tan(\varphi_{t}) = - \frac{\gamma^{-}}{\gamma^{+}},
\end{equation}

\noindent such that $\bm{\upsilon}_{t}^{\prime} = (1/ ||\bm{\upsilon}_{\textrm{B}}||) \; [u_{\textrm{B}}\tau^{\prime} \; v_{\textrm{B}}\tau^{\prime} \; w_{\textrm{B}}\tau^{\prime}]^{\textrm{T}} = [\tilde{\upsilon}^{1}_{t} \textrm{e}^{\textrm{i}\beta_{0} z} \; \tilde{\upsilon}^{2}_{t} \textrm{e}^{\textrm{i}\beta_{0} z} \; \tilde{\upsilon}^{3}_{t} \textrm{e}^{\textrm{i}\beta_{0} z}]^{\textrm{T}}$. The reader is referred to Casacuberta \textit{et al.} \cite{Casacuberta22JFM} for further details on the mathematical definition of $\bm{\upsilon}_{t}^{\prime}$.

The complementary field $\bm{\upsilon}_{n}^{\prime}$ in Eq.~(\ref{eq:pertDecompBF}) characterizes the cross-stream counterpart of $\bm{\upsilon}^{\prime}$; that is, the perturbation which, at every point in space, acts normal to base-flow streamlines. By introducing the expression of $\bm{\upsilon}_{t}^{\prime}$ (Eqs.~\ref{eq:tanDef} and \ref{componentTan}) into Eq.~(\ref{eq:pertDecompBF}), a given component of $\bm{\upsilon}_{n}^{\prime} = [\upsilon^{\prime 1}_{n} \; \upsilon^{\prime 2}_{n} \; \upsilon^{\prime 3}_{n}]^{\textrm{T}} = [\tilde{\upsilon}^{1}_{n} \textrm{e}^{\textrm{i}\beta_{0} z} \; \tilde{\upsilon}^{2}_{n} \textrm{e}^{\textrm{i}\beta_{0} z} \; \tilde{\upsilon}^{3}_{n} \textrm{e}^{\textrm{i}\beta_{0} z}]^{\textrm{T}}$ may be expressed as
\begin{equation}\label{eq:normalPert}
\upsilon_{n}^{\prime k}= \underbrace{\sqrt{|\tilde{v}^{k}|^2 + (\xi v^{k}_{\rm{B}})^2 -2 \xi v^{k}_{\rm{B}} |\tilde{v}^{k}| \cos(\varphi^{v^{k}} - \varphi_{t})}}_{|\tilde{\upsilon}^{k}_{n}|}\;\textrm{e}^{\textrm{i} (\beta_{0} z + \varphi^{\upsilon^{k}}_{n})}, \; \; k = 1,2,3,
\end{equation}
with $v^{1} = u$, $v^{2} = v$, $v^{3} = w$, $\xi = \sqrt{(\gamma^{+})^2 + (\gamma^{-})^2} / ||\bm{\upsilon}_{\rm{B}}||^2$, and associated phase
\begin{equation}\label{eq:phaseNorm}
\tan(\varphi^{\upsilon^{k}}_{n}) = \frac{|\tilde{v}^{k}| \sin(\varphi^{v^{k}}) - \xi v^{k}_{\rm{B}} \sin(\varphi_{t})}{|\tilde{v}^{k}| \cos(\varphi^{v^{k}}) -  \xi v^{k}_{\rm{B}} \cos(\varphi_{t})}.
\end{equation}

\noindent Considering a fully three-dimensional (base) flow field, the direction associated to $\bm{\upsilon}^{\prime}_{t}$ is uniquely defined \mbox{($\hat{\boldsymbol{t}} = \bm{\upsilon}_{\textrm{B}} / ||\bm{\upsilon}_{\textrm{B}}||$)}. However, the direction of $\bm{\upsilon}^{\prime}_{n}$ requires a further closure condition. In the present work, $\bm{\upsilon}^{\prime}_{n}$ is determined inherently by the difference between $\bm{\upsilon}^{\prime}$ and $\bm{\upsilon}^{\prime}_{t}$ (Eq.~\ref{eq:pertDecompBF}). It is stressed that $\bm{\upsilon}_{t}^{\prime}$ and $\bm{\upsilon}_{n}^{\prime}$ are both spanwise-periodic and complex orthogonal, i.e.\ $\bm{\upsilon}_{t}^{\prime} \bm{\cdot} \bm{\upsilon}_{n}^{\prime} = 0$ \citep{Casacuberta22JFM}, where the dot here denotes standard Hermitian inner product. As such, alike a classic wall-oriented perturbation representation, both components $\bm{\upsilon}_{t}^{\prime}$ and $\bm{\upsilon}_{n}^{\prime}$ exhibit wave-like form. 

In essence, $\bm{\upsilon}^{\prime}_{t}$ characterizes the regions of streamwise-velocity excess and deficit in the flow, whereas $\bm{\upsilon}^{\prime}_{n}$ represents typically the weak cross-stream flow pattern that is efficient in redistributing momentum. Structurally, the field $\bm{\upsilon}_{n}^{\prime}$ manifests as perturbation rolls (i.e.\ streamwise-vortical structures). As such, it is remarked that in the flow problems analyzed in this article, the streamwise-velocity perturbation component, $\bm{\upsilon}^{\prime}_{t}$, is the main contribution to the total perturbation kinetic energy,
\begin{equation}\label{eq:defE}
E_{V} = \frac{1}{2} \int_{V} \left( u^{\prime \; \dagger} u^{\prime} +  v^{\prime \; \dagger} v^{\prime} + w^{\prime \; \dagger} w^{\prime} \right) \; \textrm{d}V = \frac{1}{2} \int_{V} \left( || \bm{\upsilon}^{\prime}_{t} ||^{2} + || \bm{\upsilon}^{\prime}_{n} ||^{2} \right) \; \textrm{d}V,
\end{equation} 

\noindent with $V$ denoting a given volume in space. In this work, $V$ will be generally defined such that its length in the spanwise direction $z$ is the primary wavelength, $(2 \pi) / \beta_{0}$, of a corresponding perturbation field.

\subsection{Decomposition of the production term of the Reynolds-Orr equation}\label{sec:decomP}

Evolution equations of the kinetic perturbation energy (Eq.~\ref{eq:defE}) are typically employed to examine mechanisms of instability growth. A popular form of evolution equation is the Reynolds-Orr equation, which is specific to solenoidal flows with a localized or spatially-periodic perturbation field \cite{Schmid01}. The reader is additionally referred to Jin \textit{et al.} \cite{Jin21} for a thorough discussion on evolution equations of kinetic perturbation energy for individual modes. 

In essence, the Reynolds-Orr equation expresses that the rate of change in time of kinetic perturbation energy in a volume $V$, $\textrm{d} E_{V} / \textrm{d}t$, equals the energy exchanged between the base flow and the perturbation field, 
\begin{equation}\label{eq:production}
P_{\beta_{0}} = - \int_{V} \hat{\bm{\upsilon}}^{\prime} \bm{\cdot} \left(\hat{\bm{\upsilon}}^{\prime} \bm{\cdot} \nabla \right) \bm{\upsilon}_{\textrm{B}} \; \textrm{d}V,
\end{equation}
so-called production, and viscous dissipation, $D_{\beta_{0}}$; i.e.\ $\textrm{d} E_{V} / \textrm{d}t = P_{\beta_{0}} + D_{\beta_{0}}$. While $D_{\beta_{0}} < 0$ always, the sign of $P_{\beta_{0}}$ is informative of the \textit{sense} of energy transfer; that is, $P_{\beta_{0}} > 0$ implies that kinetic energy is transferred from the base flow to the perturbation field and vice-versa for $P_{\beta_{0}} < 0$. 

The present article analyzes the behaviour of perturbations in shear flow by inspecting the manner by which they gain or lose energy to the laminar base flow. To that end, the behaviour of $P_{\beta_{0}}$ (Eq.~\ref{eq:production}) is scrutinized by decomposing it into four contributions \cite{Albensoeder01},
\begin{equation}\label{eq:firstPDecomp}
P_{\beta_{0}} = I^{\beta_{0}}_{1} + I^{\beta_{0}}_{2} + I^{\beta_{0}}_{3} + I^{\beta_{0}}_{4},
\end{equation}

\noindent by introducing Eq.~(\ref{eq:pertDecompBF}) into Eq.~(\ref{eq:production}). Each term $I^{\beta_{0}}_{m}$ of Eq.~(\ref{eq:firstPDecomp}) may be associated to a distinct mechanism relating the exchange of kinetic energy between the base flow and the perturbation field. Namely, \mbox{$I^{\beta_{0}}_{1} = -\int_{V} \hat{\bm{\upsilon}}_{n}^{\prime} \bm{\cdot} \left( \hat{\bm{\upsilon}}_{n}^{\prime} \bm{\cdot} \nabla  \right) \bm{\upsilon}_{\textrm{B}}\;\textrm{d}V$} and \mbox{$I^{\beta_{0}}_{4} = -\int_{V} \hat{\bm{\upsilon}}_{t}^{\prime} \bm{\cdot} \left( \hat{\bm{\upsilon}}_{t}^{\prime} \bm{\cdot} \nabla  \right) \bm{\upsilon}_{\textrm{B}}\;\textrm{d}V$} represent self-induction mechanisms of respectively the cross-stream- and streamwise-velocity perturbation components. The term $I^{\beta_{0}}_{2}$ characterizes the lift-up effect \citep{Lanzerstorfer12,Picella18} and quantifies the exchange of kinetic energy between the base-flow and the streamwise-velocity perturbation ($\bm{\upsilon}^{\prime}_{t}$) by the action of the cross-stream-velocity perturbation ($\bm{\upsilon}^{\prime}_{n}$) on the base-flow shear. It reads
\begin{equation}\label{eq:I2integral}
I^{\beta_{0}}_{2} = -\int_{V} \hat{\bm{\upsilon}}_{t}^{\prime} \bm{\cdot} \left( \hat{\bm{\upsilon}}_{n}^{\prime} \bm{\cdot} \nabla  \right) \bm{\upsilon}_{\textrm{B}}\;\textrm{d}V = - \frac{2 \pi}{\beta_{0}} \int_{S} \Lambda^{\beta_{0}}_{2} \; \textrm{d}x\textrm{d}y
\end{equation}

\noindent with
\begin{equation}\label{eq:liftUp01}
\begin{split}
\Lambda^{\beta_{0}}_{2}(x,y) = &  \;\;\; \left( \tilde{\upsilon}^{1\phantom{\dagger}}_{t} \tilde{\upsilon}^{1 \dagger}_{n} + \textrm{c.c.} \right) \frac{\partial u_{\textrm{B}}}{\partial x} + \underbrace{\left( \tilde{\upsilon}^{1\phantom{\dagger}}_{t} \tilde{\upsilon}^{2 \dagger}_{n} + \textrm{c.c.} \right) \frac{\partial u_{\textrm{B}}}{\partial y}}_{\kappa^{\beta_{0}}_{2}}
+ \left( \tilde{\upsilon}^{2\phantom{\dagger}}_{t} \tilde{\upsilon}^{1\dagger}_{n} + \textrm{c.c.} \right) \frac{\partial v_{\textrm{B}}}{\partial x} \\ 
& + \left( \tilde{\upsilon}^{2\phantom{\dagger}}_{t} \tilde{\upsilon}^{2\dagger}_{n} + \textrm{c.c.} \right) \frac{\partial v_{\textrm{B}}}{\partial y}
+ \left( \tilde{\upsilon}^{3\phantom{\dagger}}_{t} \tilde{\upsilon}^{1\dagger}_{n} + \textrm{c.c.} \right) \frac{\partial w_{\textrm{B}}}{\partial x} +
\underbrace{\left( \tilde{\upsilon}^{3\phantom{\dagger}}_{t} \tilde{\upsilon}^{2\dagger}_{n} + \textrm{c.c.} \right) \frac{\partial w_{\textrm{B}}}{\partial y}}_{\delta^{\beta_{0}}_{2}},
\end{split}
\end{equation}

\noindent where $S$ denotes the $x$-$y$ cross-sectional surface of the volume $V$, and the numerical superindices of $\tilde{\upsilon}_{t}$ and $\tilde{\upsilon}_{n}$ ($1,2,3$) indicate ordering of corresponding vector components (Eq.~\ref{eq:pertDecompBF}). The role of $\kappa^{\beta_{0}}_{2}$ and $\delta^{\beta_{0}}_{2}$ in Eq.~(\ref{eq:liftUp01}) will be discussed in detail in $\S$ \ref{sec:inOutPhase}. The term \mbox{$I^{\beta_{0}}_{3} = -\int_{V} \hat{\bm{\upsilon}}_{n}^{\prime} \bm{\cdot} \left( \hat{\bm{\upsilon}}_{t}^{\prime} \bm{\cdot} \nabla  \right) \bm{\upsilon}_{\textrm{B}}\;\textrm{d}V$} characterizes the effect opposite to $I^{\beta_{0}}_{2}$, a phenomenon which has been identified in transient growth scenarios in axisymmetric vortices \citep{Antkowiak07}. It must be emphasized that the \textit{reverse lift-up effect} discussed in this present article concerns the reversal of the sign of $I^{\beta_{0}}_{2}$, and has no explicit connection with the role of $I^{\beta_{0}}_{3}$. In this regard, it is well known that the sign of each term $I^{\beta_{0}}_{m}$ stemming from the decomposition of the production term (Eq.~\ref{eq:firstPDecomp}) informs whether kinetic energy is transferred from the base flow to the perturbation field (i.e.\ the process is destabilising), $I^{\beta_{0}}_{m} > 0$, or vice-versa (i.e.\ the process is stabilising), $I^{\beta_{0}}_{m} < 0$, $m = 1$-$4$ \cite{Albensoeder01}. Finally, it is noted that a treatment analogous to Eq.~(\ref{eq:liftUp01}) may be performed for terms $I^{\beta_{0}}_{1} = (-2 \pi / \beta_{0}) \int_{S} \Lambda^{\beta_{0}}_{1} \; \textrm{d}x\textrm{d}y$, $I^{\beta_{0}}_{3} = (-2 \pi / \beta_{0}) \int_{S} \Lambda^{\beta_{0}}_{3} \; \textrm{d}x\textrm{d}y$, $I^{\beta_{0}}_{4} =  (-2 \pi / \beta_{0}) \int_{S} \Lambda^{\beta_{0}}_{4} \; \textrm{d}x\textrm{d}y$ and, by the inherent addition of the complex conjugate, Eqs.~(\ref{eq:I2integral}) and (\ref{eq:liftUp01}) are real-valued.  

\subsection{Description and set-up of flow problems}

The main analysis of this article focuses on the mechanisms of stationary CFI stabilization by a forward-facing-step ($\S$ \ref{sec:resultStep}). First, however, two model problems are discussed to illustrate main concepts in simpler flow environments. The corresponding flow problems and the numerical set-up are introduced and described next. Results will be discussed later on in $\S$ \ref{sec:results}. 

\subsubsection{Model problem I: plane Poiseuille flow}\label{sec:setUpChannel}

\begin{figure}
\begin{center}
\includegraphics{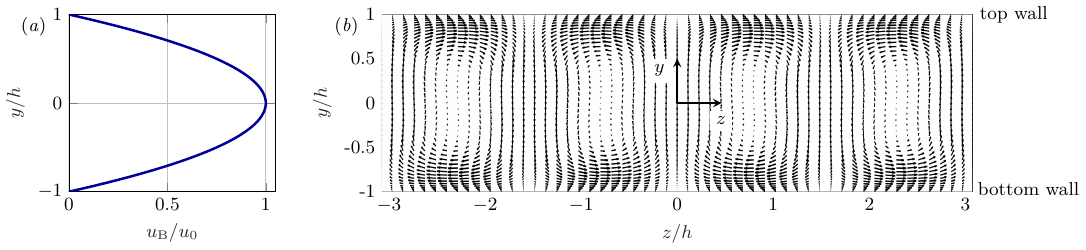}
\caption{\label{fig:channelSetupPlot} {Sketch of model problem I: base-flow profile $u_{\textrm{B}}$ (\textit{a}) and organization of \textit{optimal} perturbation rolls $\hat{\bm{\upsilon}}^{\prime} = [0 \; \hat{v}^{\prime} \; \hat{w}^{\prime}]^{\textrm{T}}$ at the initial time instant (\textit{b}).}}
\end{center}
\end{figure}

The first model problem entails \textit{optimal} perturbations in incompressible plane Poiseuille flow. It represents a classic example of perturbation growth driven by the lift-up effect \cite{Schmid01}, thus it is an appealing case to illustrate the concept of a reverse lift-up effect. Furthermore, the simple topology of the unperturbed base flow yields a straightforward representation of underlying equations. Namely, the base flow is invariant in the streamwise, $x$, and in the spanwise, $z$, directions and the profile of the $x$-velocity, $u_{\textrm{B}}$, is parabolic along the wall-normal direction, $y$. The base-flow field hence reads $\bm{v}_{\textrm{B}} = [u_{\textrm{B}}(y) \; 0 \; 0]^{\textrm{T}}$ with $u_{\textrm{B}} = u_{0} \left(1 - y^2/h^2 \right)$ and $u_{0}$ denoting the peak velocity of $u_{\textrm{B}}$ at centerline; i.e.\ at $y = 0$. The walls are placed at $y = \pm h$. A sketch of the flow problem and coordinate system is depicted in Fig.~\ref{fig:channelSetupPlot}.

The aim of this model problem is to illustrate the role of the lift-up effect in the growth and decay of \textit{optimal} perturbations \cite{Schmid14}. The flow problem entails $x$-invariant counter-rotating perturbation rolls prescribed as an initial perturbation condition. These perturbation rolls are flow patterns of wave-like form that act orthogonal to the base-flow streamlines; i.e.\ they represent a cross-stream perturbation as characterized by $\bm{\upsilon}_{n}^{\prime}$ in Eq.~(\ref{eq:pertDecompBF}). By acting on the base-flow shear, $\partial u_{\textrm{B}} / \partial y$, the rolls redistribute base-flow momentum and induce streamwise perturbation streaks (i.e.\ regions of streamwise-velocity deficit and excess as characterized by $\bm{\upsilon}_{t}^{\prime}$ in Eq.~(\ref{eq:pertDecompBF}) through the lift-up effect \cite{Schmid01}. Alike the perturbation rolls, the streamwise streaks are invariant in the $x$-direction. While the streaks display a rapid initial growth in time, the perturbation rolls do not change significantly their topology as time evolves. These \textit{optimal} initial perturbation rolls, i.e.\ the flow pattern which yields the largest transient growth of kinetic perturbation energy, and the corresponding optimal perturbation response for a chosen time horizon ($t u_{0} / h = 27.895$) have been computed with the code $\textsf{OptimalDisturbance.m}$ provided by Schmid and Brandt \cite{Schmid14}. Particularly, we choose a spanwise perturbation wavenumber of $\beta_{0} h = 2.04$ and $Re = u_{0} h / \nu = 1000$ with $\nu$ denoting kinematic viscosity. 

\subsubsection{Model problem II: Blowing-Suction in two-dimensional boundary-layer flow}\label{sec:setUpBS}

A second model problem is discussed to illustrate the principle of a reverse lift-up effect in the context of boundary-layer flow. This model problem entails streamwise perturbation streaks in a two-dimensional spatially-accelerating (i.e.\ favorable pressure gradient) boundary layer over a flat plate. When developing in the streamwise direction, the streaks interact with localized steady wall Blowing-Suction (BS). The elements of the perturbation field in this (model) flow problem resemble structurally those of the main case in this work ($\S$ \ref{sec:resultStep}). As such, the multiple similarities between both perturbation scenarios paves the road for full characterization of the reverse lift-up effect in the highly deformed three-dimensional step flow discussed in $\S$ \ref{sec:resultStep}. Moreover, for the sake of representation, major features of the set-up of this model problem are similar to the main case ($\S$ \ref{ref:flowProblem}). Namely, the acceleration of the free-stream in the streamwise direction, $x$, and the inlet Reynolds number. However, it is emphasized that the base flow is purely two-dimensional and unswept in the present model problem; i.e.\ $\bm{v}_{\textrm{B}} = [u_{\textrm{B}}(x,y) \; v_{\textrm{B}}(x,y) \; 0]^{\textrm{T}}$. 

As such, this model problem is configured to simulate the interaction of two distinct sets of perturbations, represented as $\hat{\bm{\upsilon}}^{\prime}$: the field $\hat{\bm{\upsilon}}^{\prime}$ is composed of a pair of (cross-stream) counter-rotating perturbation rolls prescribed at the inflow and streamwise perturbation streaks which amplify in $x$ as a result of the lift-up effect \cite{Luchini00} induced by the action of the rolls on the base-flow shear. The wavelength in the spanwise direction, $z$, of the perturbation rolls is identically that of the CFI mode in the main case of this article ($\S$ \ref{ref:flowProblem}) and equals the domain length in the spanwise direction. The dimensions of the computational domain are $0 \leq x / \delta_{0} \leq 123$ in the streamwise direction, $0 \leq y / \delta_{0} \leq 26$ in the wall-normal direction, and $-4.86 \leq z / \delta_{0} \leq 4.86$ in the spanwise direction. Here, $\delta_{0}$ denotes the $99 \%$ boundary-layer thickness at inflow. This parameter is employed for non-dimensionalization purposes, together with $u_{\infty}$, which denotes the inflow free-stream velocity. It is emphasized that both the pre-imposed inflow rolls and the spatially-forming streamwise streaks are stationary perturbation structures; i.e.\ $\partial \hat{\bm{\upsilon}}^{\prime} / \partial t = 0$.

Corresponding flow fields of this model problem are computed numerically in a sequential manner. First, the unperturbed base flow is computed by performing Direct Numerical Simulations (DNS) with the conservative finite-volume solver INCA; see Casacuberta \textit{et al.} \cite{Casacuberta22JFM} for details on the numerical discretization of equations and the class of boundary conditions employed.  Second, steady-state DNS is performed to obtain the streamwise streaks in the boundary layer. In this second simulation, the inflow boundary condition includes the pair of cross-stream rolls superimposed on the laminar base-flow profile. A fully steady-state DNS solution is enforced by use of the Selective Frequency Damping (SFD) method \cite{Akervik06,Casacuberta18}. Finally, in a third simulation, steady Blowing-Suction (BS) at the wall is applied. To that end, a BS strip is introduced, in which the wall-normal velocity at the wall is modulated harmonically as follows:
\begin{equation}\label{eq:BSFormula}
v_{\textrm{BS}}(x,0,z) = f_{s}(x) A_{\textrm{BS}} \cos(\beta_{0}z + \phi_{\textrm{BS}}),
\end{equation}

\noindent where $\beta_{0}$ denotes the perturbation wavenumber in $z$, $A_{\textrm{BS}}$ is the amplitude of the wall-normal-velocity modulation, and $f_{s}$ is a smooth function which establishes a gradual evolution of the BS velocity between the starting (\mbox{$x_{\textrm{BS},\textrm{start}} = 57.62 \delta_{0}$}) and ending (\mbox{$x_{\textrm{BS},\textrm{end}} = 62.62 \delta_{0}$}) positions of the BS strip:
\begin{equation}
f_{s} = \left( \frac{4 \left( x - x_{\textrm{BS},\textrm{start}} \right) \left( x_{\textrm{BS},\textrm{end}} - x \right)}{(x_{\textrm{BS},\textrm{end}} - x_{\textrm{BS},\textrm{start}})^2} \right)^3.
\end{equation}

\noindent The width of the strip in the $x$-direction is kept small to produce a \textit{local} effect, which is representative of the flow environment around the step discussed in $\S$ \ref{sec:resultStep}. The spanwise phase of the wall-normal velocity in the strip, $\phi_{\textrm{BS}} = \pi / 2$ (Eq.~\ref{eq:BSFormula}), is chosen such that the BS strip acts by locally stabilizing the incoming streamwise streaks. This is achieved by aligning the blowing region (i.e.\ $v_{\textrm{BS}} > 0$) with the incoming high-speed streak (i.e.\ $u^{\prime} > 0$) and the suction region (i.e.\ $v_{\textrm{BS}} < 0$) with the incoming low-speed streak (i.e.\ $u^{\prime} < 0$). The amplitude of $v_{\textrm{BS}}$, $A_{\textrm{BS}} = 1 \times 10^{-6} u_{\infty}$, yields a scenario of linearly dominated perturbation evolution. To retrieve perturbation information, the full perturbation field is decomposed into the sum of spanwise Fourier modes. The reader is referred to the next section (\ref{ref:flowProblem}) for details on the latter. 

\subsubsection{Step in three-dimensional swept-wing flow}\label{ref:flowProblem}

The main flow problem of this work considers the interaction between stationary CFI and forward-facing step in three-dimensional swept-wing flow. We model the swept-wing flow as incompressible flat-plate flow (i.e.\ neglecting wall curvature effects) with a prescribed favourable pressure gradient in the direction of the wing chord. The set-up of the present DNS is largely similar to that employed recently by Casacuberta \textit{et al.} \cite{Casacuberta22JFM}, which in turn has been guided by experimental work \cite{Rius21} on a $45^{\circ}$ swept wing. In particular, $Re = u_{\infty} \delta_{0} / \nu = 791.37$ with $u_{\infty} = 15.10$ m/s denoting the free-stream chordwise velocity and $\delta_{0} = 7.71 \times 10^{-4}$~m is the $99 \%$ boundary-layer height relative to $u_{\infty}$. All reference quantities are taken at the DNS inflow, which is placed virtually at $5\%$ of the reference wing chord. 

\begin{figure}
\begin{center}
\includegraphics{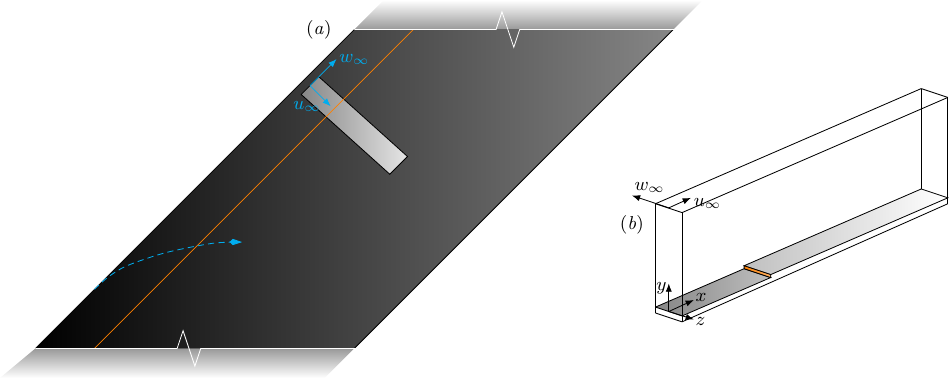}
\caption{\label{fig:sketchStep} {Sketch of the step-flow problem: top view of a reference (virtually infinite) swept wing (dark gray), computational domain (light gray), distributed step (solid orange line), trajectory of inviscid streamline (dashed cyan line), and decomposition of the free-stream velocity vector at inlet (\textit{a}). Computational domain and coordinate system (\textit{b}).}}
\end{center}
\end{figure}

The main coordinate system, $\bm{x} = [x \; y \; z]^{\textrm{T}}$, is oriented relative to the wing, i.e.\ $x$ is aligned with the leading-edge-orthogonal (or, chordwise) direction, $z$ is aligned with the leading-edge-parallel (or, spanwise) direction, and $y$ points in the direction normal to the wall. The velocity vector reads $\bm{v} = [u \; v \; w]^{\textrm{T}}$ with $u, v,$ and $w$ respectively denoting velocity components in the chordwise, wall-normal, and spanwise directions. In this flow problem, the decomposition of the velocity field into the three-dimensional base flow, $\bm{v}_{\textrm{B}}$ with $\partial \bm{v}_{\textrm{B}} / \partial z = 0$, and a three-dimensional stationary perturbation field, $\hat{\bm{v}}^{\prime}$, reads $\bm{v}(x,y,z) = \bm{v}_{\textrm{B}}(x,y) + \hat{\bm{v}}^{\prime}(x,y,z)$.

The perturbation field entails a main (fundamental) perturbation component (i.e.\ primary wavelength) as well as high-order harmonic (i.e.\ smaller wavelength) perturbation components. As such, capitalizing on the periodic nature of the velocity-perturbation field in the spanwise direction, it is instructive to decompose further $\hat{\bm{\upsilon}}^{\prime}$ into spanwise Fourier modes, i.e.
\begin{equation}\label{eq:FourierAnsantz}
\hat{\bm{\upsilon}}^{\prime} = \sum_{j = -N}^{N} \tilde{\bm{\upsilon}}_{j}(x,y) \ \textrm{e}^{ \textrm{i} j \beta_{0} z},
\end{equation}

\noindent where $\tilde{\bm{\upsilon}}_{j} \in \mathbb{C}^{3}$ are the Fourier coefficients, $N$ is the number of modes, $\beta_{0}$ is the fundamental spanwise wavenumber, and $\textrm{i}^2 = -1$. It is noted that $\tilde{\bm{\upsilon}}_{-j} = \tilde{\bm{\upsilon}}^{\dagger}_{j}$. Primary-wavelength ($j = 1$) perturbation effects are the main scope of the present analysis. Henceforth, only the primary-wavelength Fourier space is considered for the remainder of the analysis and, for simplicity, $\bm{\upsilon}^{\prime} = \tilde{\bm{\upsilon}}_{1} \ \textrm{e}^{ \textrm{i} \beta_{0} z} = [u^{\prime} \; v^{\prime} \; w^{\prime}]^{\textrm{T}}$ will be hereafter referred to as \textit{the perturbation field}. Thus, $\hat{\bm{\upsilon}}^{\prime}={\bm{\upsilon}}^{\prime}+{\bm{\upsilon}}^{\prime\dagger}$. It has been shown that, for the present choice of inflow CFI amplitude, the evolution of the primary-wavelength perturbation is essentially linearly-dominated near the step \cite{Casacuberta22JFM}. However, it is stressed that the present DNS grid resolution is sufficient to accurately resolve the evolution of higher-order crossflow harmonics; see Casacuberta \textit{et al.} \cite{Casacuberta22JFM} in this regard. Following the nomenclature introduced in $\S$ \ref{sec:mainFormulation}, the modulus of the components of the Fourier coefficient $\tilde{\bm{\upsilon}}_{1}$ (Eq.~\ref{eq:FourierAnsantz}) read $|\tilde{u}|$, $|\tilde{v}|$, and $|\tilde{w}|$ (in a Cartesian wall-oriented representation) and are referred to as amplitude functions; a corresponding perturbation growth rate in $x$ is evaluated as
\begin{equation}\label{eq:growthRate}
\alpha^{q}_{i} = - \frac{1}{A^{q}} \frac{\textrm{d} A^{q}}{\textrm{d}x},
\end{equation}

\noindent where $A^{q} = A^{q}(x)$ is the amplitude of a velocity component $q$ (e.g. the chordwise-velocity perturbation $u$). 

The forward-facing step, which is homogeneous along the spanwise direction, is placed at $20 \%$ of the chord of the wing model, which corresponds to $x = 177.62 \delta_{0}$ in the DNS set-up. For the sake of clarity, the coordinate $x_{\textrm{st}} = x - 177.62\delta_{0}$ indicating the distance relative to the step will be employed additionally. The main step height analysed in this work, $h / \delta_{0} = 0.97$, which was found to yield stabilization in the reference numerical work \cite{Casacuberta22JFM}, corresponds to approximately $50 \%$ of the undisturbed boundary-layer height at the virtual step location. To analyze step-height effects, two additional step geometries are considered in the present study, namely $h / \delta_{0} = 0.59$ and $0.76$. Periodic boundary conditions are prescribed at the transverse boundaries and the spanwise domain length is set equal to the fundamental crossflow wavelength, thus allowing growth of perturbations exclusively in $x$. Furthermore, the Selective Frequency Damping (SFD) method \citep{Akervik06,Casacuberta18} is applied to ensure the fully stationary nature of the developed flow and to constrain the analysis to stationary mechanisms. 

\section{Analysis of model problems}\label{sec:results}

\subsection{Model problem I: plane Poiseuille flow}\label{sec:modelI}

\begin{figure}
\begin{center}
\includegraphics{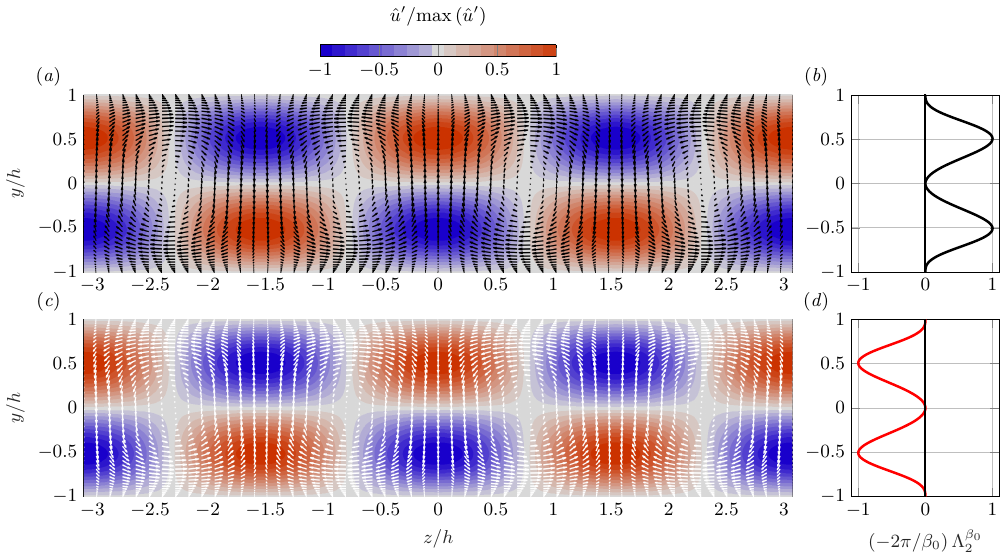}
\caption{\label{fig:channelPlot} {Organization of perturbations in a \textit{classic} (\textit{a}) and a \textit{reverse} (\textit{c}) scenario of the lift-up effect in plane Poiseuille flow at a fixed time instant: cross-stream rolls (arrows) and streamwise streaks (color contour). Integrand of the lift-up term $I^{\beta_{0}}_{2}$ normalized with respect to its maximum value in $y$, characterizing the top (\textit{b}) and bottom (\textit{d}) cases.}}
\end{center}
\end{figure}

The first model problem, plane Poiseuille flow, is a canonical example of perturbation amplification by the classic lift-up effect \cite{Schmid01}. This flow problem entails streamwise-invariant perturbation rolls distributed periodically in the spanwise direction, $z$, which are prescribed in the form of an initial perturbation condition (see the set-up description in $\S$ \ref{sec:setUpChannel}). As time evolves, streamwise-invariant velocity-perturbation streaks form. This is illustrated in Fig.~\ref{fig:channelPlot}(\textit{a}) portraying a $y$-$z$ plane of the perturbation response at the chosen time horizon ($t u_{0} / h = 27.895$) obtained through a singular value decomposition of the corresponding matrix exponential \cite{Schmid14}.

We recall the Reynolds-Orr equation relating the rate of change of kinetic perturbation energy, $E_{V}$ (Eq.~\ref{eq:defE}), in time with the effect of production, $P_{\beta_{0}}$, and viscous dissipation, $D_{\beta_{0}}$; see $\S$ \ref{sec:decomP}. The pre-imposed perturbation rolls act by redistributing low- and high-momentum fluid and thus they feed growth to streamwise streaks; i.e.\ production $P_{\beta_{0}} > 0$, naturally implying that the streamwise streaks grow by extracting energy from the base-flow shear. Concerning the overarching discussion in this article, the realization that the lift-up effect --characterized by $I^{\beta_{0}}_{2}$ ($\S$ \ref{sec:decomP})-- is the main mechanism driving here the perturbation amplification \cite{Schmid01} follows from the fact that $I^{\beta_{0}}_{1} = I^{\beta_{0}}_{3} = I^{\beta_{0}}_{4} = 0 \Rightarrow P_{\beta_{0}} = I^{\beta_{0}}_{2}$ (Eq.~\ref{eq:firstPDecomp}). This is illustrated as follows: the base flow is a parallel flow (i.e.\ $\partial u_{\textrm{B}} / \partial y$ is the only active base-flow shear), thus the integrands of $I^{\beta_{0}}_{m}, m = 1$-$4$ (Eqs. \ref{eq:I2integral} and \ref{eq:liftUp01}), simplify as
\begin{align} \label{eq:I12}\tag{3.2,\textit{a}}
& \Lambda^{\beta_{0}}_{1} = \tilde{\upsilon}^{1\phantom{\dagger}}_{n} \tilde{\upsilon}^{2 \dagger}_{n} \; \frac{\partial u_{\textrm{B}}}{\partial y} + \textrm{c.c.}, 
& \Lambda^{\beta_{0}}_{2} = &  \; \tilde{\upsilon}^{1\phantom{\dagger}}_{t} \tilde{\upsilon}^{2 \dagger}_{n} \; \frac{\partial u_{\textrm{B}}}{\partial y} + \textrm{c.c.}, 
\\ \nonumber 
& \Lambda^{\beta_{0}}_{3} = \tilde{\upsilon}^{1\phantom{\dagger}}_{n} \tilde{\upsilon}^{2 \dagger}_{t} \; \frac{\partial u_{\textrm{B}}}{\partial y} + \textrm{c.c.}, 
& \Lambda^{\beta_{0}}_{4} = &  \; \tilde{\upsilon}^{1\phantom{\dagger}}_{t} \tilde{\upsilon}^{2 \dagger}_{t} \; \frac{\partial u_{\textrm{B}}}{\partial y} + \textrm{c.c.} \label{eq:I34}\tag{3.2,\textit{b}}
\end{align}

\noindent In turn, particularly for the present flow problem, the perturbation components tangential and normal to base-flow streamlines may be related explicitly to the perturbation components tangential and normal to the wall as $\bm{\upsilon}^{\prime}_{t} = [u^{\prime} \; 0 \; 0]^{\textrm{T}}$ and $\bm{\upsilon}^{\prime}_{n} = [0 \; v^{\prime} \; w^{\prime}]^{\textrm{T}}$. Therefore, $\Lambda^{\beta_{0}}_{1} = \Lambda^{\beta_{0}}_{3} = \Lambda^{\beta_{0}}_{4} = 0$ and 
\begin{equation}\label{eq:prodChannel}
P_{\beta_{0}} = I^{\beta_{0}}_{2} = - \frac{2 \pi}{\beta_{0}} \int_{S}  ( \tilde{u} \; \tilde{v}^{\dagger} \; \frac{\partial u_{\textrm{B}}}{\partial y} + \textrm{c.c.} ) \; \textrm{d}S = - \frac{4 \pi}{\beta_{0}} \int_{S} |\tilde{u}|  |\tilde{v}|  \frac{\partial u_{\textrm{B}}}{\partial y} \cos(\varphi^{u} - \varphi^{v}) \; \textrm{d}S.
\end{equation}

\noindent Equation (\ref{eq:prodChannel}) highlights that the \textit{sense} of energy transfer between the base flow and the perturbation field (through the lift-up effect) is dictated by the sign of $\cos(\varphi^{u} - \varphi^{v})$; i.e.\ by the phase difference between cross-stream rolls and streamwise streaks establishing their relative placement along the spanwise direction $z$. By their relative phase in the present configuration, see Fig.~\ref{fig:channelPlot}(\textit{a}), $(-2 \pi / \beta_{0} ) \; \Lambda^{\beta_{0}}_{2} > 0$ for all $y$ (Fig.~\ref{fig:channelPlot}(\textit{b})). The latter implies that the action of the cross-stream rolls ($\bm{\upsilon}^{\prime}_{n}$) acts destabilizing and base-flow kinetic energy feeds growth to the streamwise streaks ($\bm{\upsilon}^{\prime}_{t}$). This illustrates the typical scenario of the classic lift-up effect. It is noted that in the present model problem, $\Lambda^{\beta_{0}}_{2} = \Lambda^{\beta_{0}}_{2}(y)$ since both the base flow and the perturbation field are invariant in $x$. 

In essence, the core analysis of this article revolves around the fact that the same principle holds, but operates in a reverse fashion, if the relative spatial placement between cross-stream rolls and streamwise streaks (i.e.\ their spanwise phase) is altered. This is illustrated as follows: consider that at the time instant depicted in Fig.~\ref{fig:channelPlot}(\textit{a},\textit{b}), the spanwise phase of the rolls, $\varphi^{v}$, is shifted by $\pi$ radians such that the term $\cos (\varphi^{u} - \varphi^{v})$ in Eq.~(\ref{eq:prodChannel}) reverts its sign. In such new perturbation environment, depicted in Fig.~\ref{fig:channelPlot}(\textit{c}), $(-2 \pi / \beta_{0} ) \; \Lambda^{\beta_{0}}_{2} < 0$ for all $y$ (Fig.~\ref{fig:channelPlot}(\textit{d})). In the new scenario, $I^{\beta_{0}}_{2} < 0$ implying that $P_{\beta_{0}} < 0$ and since $D_{\beta_{0}} < 0$ always, $\textrm{d}E_{V} / \textrm{d}t < 0$. Therefore, the perturbation field undergoes stabilization locally in time, which shall be interpreted as the flow exhibiting a tendency towards recovering the original (unperturbed) laminar base state: low-momentum fluid is displaced towards the regions of streamwise-velocity excess (i.e.\ red regions in Fig. \ref{fig:channelPlot}) and high-momentum fluid is displaced towards the regions of streamwise-velocity deficit (i.e.\ blue regions in Fig. \ref{fig:channelPlot}). Consequently, a \textit{reverse lift-up effect} now takes place. At present, we have exemplified such a reverse lift-up effect on a pre-existing streaky flow field by altering artificially the spatial organization of the perturbation content acting normal (the rolls) and tangential (the streaks) to the base flow. In $\S$ \ref{sec:resultStep} it will be shown that essentially the same perturbation effect, but conditioned naturally by an abrupt spatial variation of the flow organization, stabilizes significantly a pre-existing convective instability. 

On a historical note, the notion that the lift-up effect is a powerful destabilizing flow mechanism originates mainly from the work of Ellingsen and Palm \cite{Ellingsen75} and Landahl \cite{Landahl75,Landahl80}. They formalized that a three-dimensional cross-stream perturbation in shear flow may induce growth of perturbation kinetic energy (by the lift-up effect) irrespective of whether the flow supports a modal (exponential) instability. While the concept of a stabilizing (reverse) lift-up effect may seem paradoxical at first glance, it actually follows naturally from the model of Ellingsen and Palm \cite{Ellingsen75} if a non-zero initial perturbation streak field is considered. This is elaborated upon in detail in Appendix A. 

\subsection{Model problem II: Blowing-Suction in two-dimensional boundary-layer flow}\label{sec:modelII}

\begin{figure}
\begin{center}
\includegraphics{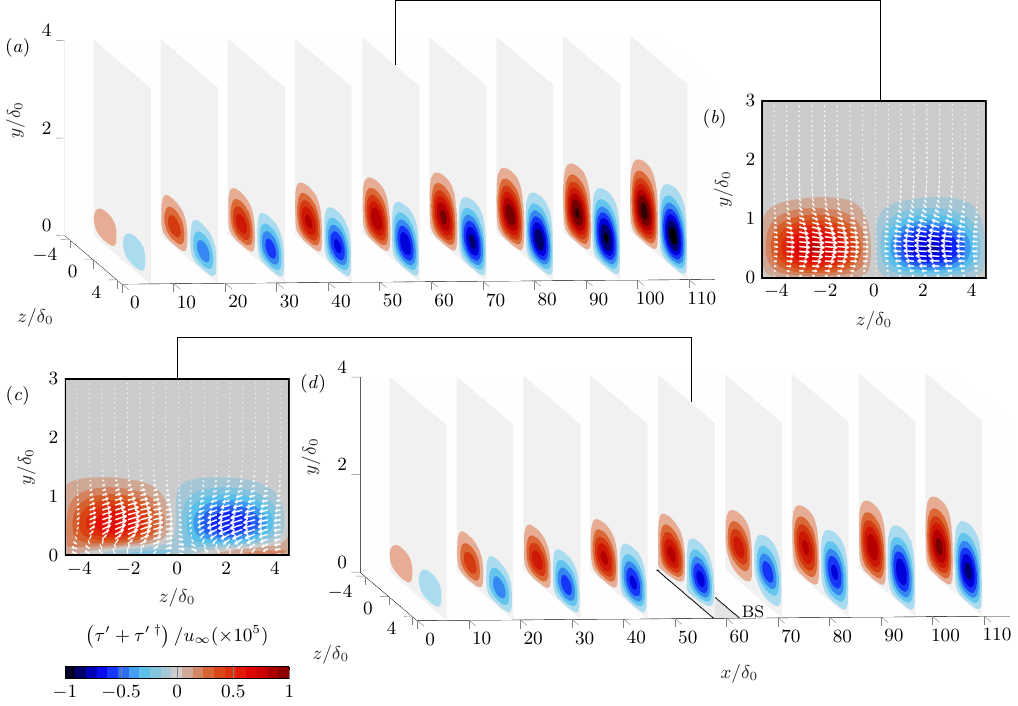}
\caption{\label{fig:blowsuc3D} {Evolution of the streamwise-velocity perturbation in no-BS(\textit{a},\textit{b}) and BS (\textit{c},\textit{d}) cases (color map): three-dimensional organization(\textit{a},\textit{d}) and $y$-$z$ planes (\textit{b},\textit{c}) at $x / \delta_{0} = 59$ with white arrows illustrating the in-plane organization of the cross-stream-velocity perturbation. Surface BS strip depicted as gray rectangle in (\textit{d}).}}
\end{center}
\end{figure}  

\begin{figure}
\begin{center}
\includegraphics{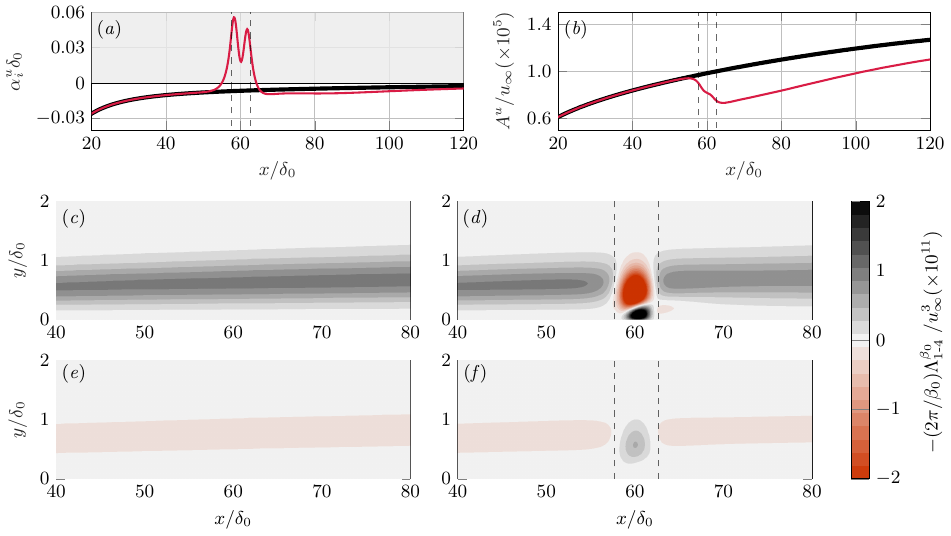}
\caption{\label{fig:BSPlotGR} {Evolution in $x$ of perturbation spatial growth rate (\textit{a}) and amplitude (\textit{b}) in reference (thick black) and BS \mbox{(thin red)} cases. Spatial organization of the integrand of $I^{\beta_{0}}_{2}$ (\textit{c},\textit{d}) and $I^{\beta_{0}}_{1} + I^{\beta_{0}}_{3} + I^{\beta_{0}}_{4}$ (\textit{e},\textit{f}) in reference (\textit{c},\textit{e}) and BS (\textit{d},\textit{f}) cases. Dashed vertical lines indicate the starting and ending positions of BS.}}
\end{center}
\end{figure}

A second model problem illustrates the concept of a reverse lift-up effect in the context of a spatially-developing boundary layer. The set-up of the flow problem includes a pair of steady counter-rotating perturbation rolls prescribed at inflow ($x = 0$). By the action of the rolls on the two-dimensional boundary layer developing over a flat plate, streamwise streaks amplify spatially in the direction of the free-stream, $x$. The streaks manifest structurally as regions of streamwise-velocity deficit and excess modulated along the spanwise direction, $z$. The spatial evolution of the streaks is illustrated in Fig.~\ref{fig:blowsuc3D}(\textit{a}) depicting the streamwise-velocity perturbation, $\tau^{\prime}$ (Eq.~\ref{eq:tanDef}). At present, the streaks grow monotonically in $x$, as highlighted by the trend of their spatial growth rate, $\alpha_{i}$, in Fig.~\ref{fig:BSPlotGR}(\textit{a}) (solid black line); it is emphasized that $\alpha_{i} < 0$ here implies growth in space. Furthermore, it is noted that, by the organization of the flow in the present model problem, the representation of velocity perturbations acting tangential to the wall in $x$, $u^{\prime}$, and tangential to the base-flow streamlines, $\tau^{\prime}$, are largely similar. 

It is well known that the (classic) lift-up effect drives the perturbation amplification in the aforementioned scenario. For instance, Luchini \cite{Luchini00} describes that ``perturbations produced in this way are driven by the lift-up phenomenon, that is by the continued accumulation over downstream distance of longitudinal-velocity differences arising from slow convection in the transverse plane. This mechanism tends to give these perturbations their characteristic aspect of elongated streaks.'' Thus, the term $I_{2}^{\beta_{0}}$ characterizing the lift-up effect naturally adds here the main contribution to energy production (Fig.~\ref{fig:BSPlotGR}(\textit{c})) in relation to negligible contributions by $I_{1}^{\beta_{0}}, I_{3}^{\beta_{0}}, I_{4}^{\beta_{0}}$ (Fig.~\ref{fig:BSPlotGR}(\textit{e})) and $I_{2}^{\beta_{0}} > 0$ for all $x$. The latter implies that the lift-up effect acts destabilizing in the present flow and kinetic energy of the base flow feeds growth to the streamwise streaks by the action of the cross-stream rolls.  

A popular method used to stabilize elongated streamwise streaks is wall Blowing-Suction; see for instance Lundell \textit{et al.} \cite{Lundell03}. To produce a stabilizing effect, a blowing region is placed underneath the high-speed streak and a suction region is placed underneath the low-speed streak. The aim of this section is to highlight that the concept of a reverse lift-up effect offers a simple way to understand and quantify the mechanism of perturbation stabilization by BS. To that end, a BS surface strip is next positioned in the region $57.62 \leq x / \delta_{0} \leq 62.62$, see Fig.~\ref{fig:blowsuc3D}(\textit{d}) illustrating the surface strip in relation to the spatially-developing streaks. In the surface strip, the wall-normal velocity it set to behave harmonically in $z$ and to \textit{oppose} the incoming streaks (see $\S$ \ref{sec:setUpBS} for details on the BS set-up). The BS thus reduces the \textit{strength} of the incoming streak system locally in space: this effect is quantified by the large increase and change in sign of the spatial growth rate, see solid red line in Fig.~\ref{fig:BSPlotGR}(\textit{a}), and corresponding decay of perturbation amplitude, see solid red line Fig.~\ref{fig:BSPlotGR}(\textit{b}). Finally, the results in Fig.~\ref{fig:BSPlotGR}(\textit{a},\textit{b}) show additionally that the original perturbation mechanism (i.e.\ the classic lift-up effect) is gradually recovered downstream of the BS strip.  

In the BS scenario, the velocity induced at the wall reverses locally the interplay between cross-stream- (white arrows in Fig.~\ref{fig:blowsuc3D}(\textit{b},\textit{c})) and streamwise- (color map in Fig.~\ref{fig:blowsuc3D}) velocity perturbations due to their relative spanwise phase. That is, the cross-stream- and streamwise-velocity perturbations act \textit{in-phase} (i.e.\ against) in the BS case, whilst they act \textit{out-of-phase} (i.e.\ in favor) in the no-BS case; see the comparison between Fig.~\ref{fig:blowsuc3D}(\textit{c}) and Fig.~\ref{fig:blowsuc3D}(\textit{b}). Following the rationale discussed in $\S$ \ref{sec:modelI}, but now in the context of spatially-developing perturbations, a reversal of the sign of $I_{2}^{\beta_{0}}$ is consequently monitored locally around the surface BS strip (see red region in Fig.~\ref{fig:BSPlotGR}(\textit{d})). From the viewpoint of production, the latter shall be interpreted as the cross-stream perturbations now acting by transferring kinetic energy of the pre-existing streamwise streaks towards the underlying flow. At the same time, Fig.~\ref{fig:BSPlotGR}(\textit{f}) shows that the mechanisms of $I_{1}^{\beta_{0}}$, $I_{3}^{\beta_{0}}$, $I_{4}^{\beta_{0}}$ remain negligible in the region of BS. Therefore, the stabilization via BS originates purely from a reverse lift-up effect; i.e.\ a reversal of the sense of kinetic-energy transfer with respect to the original (classic lift-up) mechanism.

\section{Classic and reverse lift-up effects in interaction between steps and CFI \label{sec:resultStep}}

The remainder of this work addresses the interaction between stationary crossflow instability and forward-facing step using the problem set-up introduced in $\S$ \ref{ref:flowProblem}. Following the discussion of model problems I ($\S$ \ref{sec:modelI}) and II \mbox{($\S$ \ref{sec:modelII})}, emphasis is placed here on the role played by the lift-up effect in altering the evolution of a pre-existing (i.e.\ incoming) crossflow instability around the step.

\subsection{Perturbation behavior at the step}\label{sec:overviewStep}

\begin{figure}
\begin{center}
\includegraphics{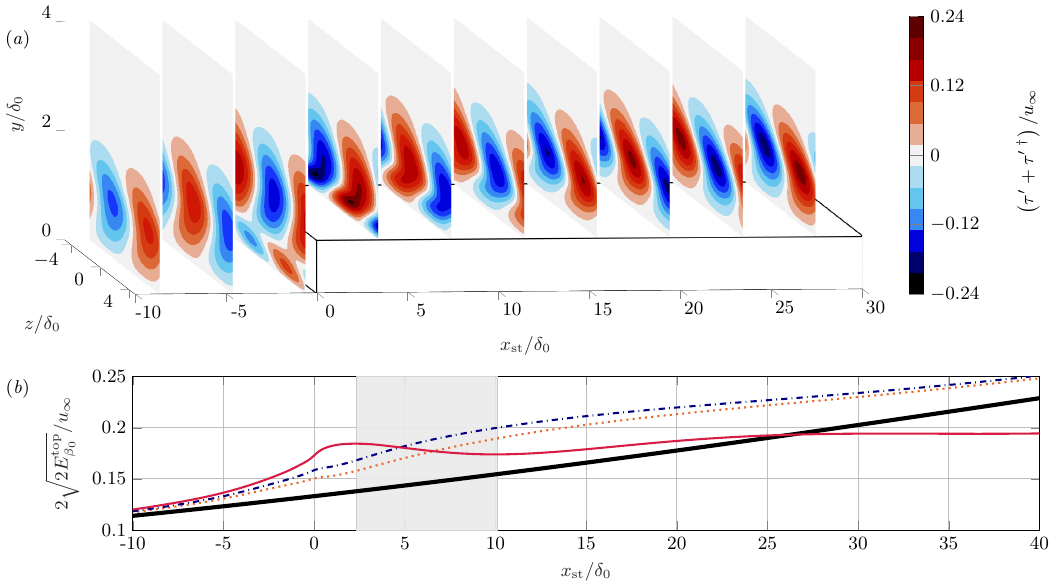}
\caption{\label{fig:evol3D} {Organization of the streamwise-velocity perturbation component around the step (\textit{a}). Evolution in $x$ of kinetic perturbation energy, $E^{\textrm{top}}_{\beta_{0}}$ (Eq.~\ref{eq:energy}), in the no-step (thick solid black) and step of $h / \delta_{0} = 0.97$ (thin solid red) cases; grey area highlights $\textrm{d} E^{\textrm{top}}_{\beta_{0}} / \textrm{d}x < 0$.  Step case of $h / \delta_{0} = 0.59$ (dotted orange line) and step case of $h / \delta_{0} = 0.76$ (dash-dotted blue) (\textit{b}).}}
\end{center}
\end{figure}

Figure \ref{fig:evol3D}(\textit{a}) illustrates the organization of the perturbation field around the step. Particularly, Fig.~\ref{fig:evol3D}(\textit{a}) depicts the streamwise-velocity perturbation component, characterized as $\tau^{\prime}$ (Eq.~\ref{eq:tanDef}), which represents approximately $98 \%$ of the total kinetic energy budget (in relation to the cross-stream component). Two distinct families of perturbation structures are recognized in the near-step regime. Namely, the pre-existing (incoming) crossflow instability, which approaches the step in the chordwise direction $x$, passes over the step, and develops further downstream of it \cite{Tufts17,Eppink20,Rius21,Casacuberta22JFM}. At the same time, newly-formed stationary velocity-perturbation streaks are induced around the step as a by-product of the interaction between the incoming instability and the step. The origin of these newly-formed streaks is currently debated \cite{Tufts17,Eppink20,Casacuberta21,Casacuberta22JFM}. Similar regions of streamwise-velocity deficit and excess arising close to the wall, just downstream of the step, have also been reported by Lanzerstorfer and Kuhlmann \cite{Lanzerstorfer12} for unswept channel forward-facing-step flow. The near-wall streaks are accompanied by corresponding (near-wall) perturbation rolls manifesting in the field $\bm{\upsilon}^{\prime}_{n}$. In previous numerical investigations, such additional near-wall stationary perturbation rolls were identified as regions of enhanced perturbation vorticity around the step \cite{Casacuberta21}. 

Casacuberta \textit{et al.} \cite{Casacuberta22JFM} note that, for the choice of step height discussed in the present article, the incoming crossflow perturbation is stabilized locally downstream of the step, when compared to reference no-step conditions. This is here quantified in Fig.~\ref{fig:evol3D}(\textit{b}) portraying the chordwise evolution of the perturbation kinetic energy evaluated as
\begin{equation}\label{eq:energy}
E^{\textrm{top}}_{\beta_{0}}(x) = \frac{1}{2} \left( u^{\prime \; \dagger} u^{\prime} +  v^{\prime \; \dagger} v^{\prime} + w^{\prime \; \dagger} w^{\prime} \right)\Big|_{\textrm{top}} = \frac{1}{2} \left( || \bm{\upsilon}^{\prime}_{t} ||^{2} + || \bm{\upsilon}^{\prime}_{n} ||^{2} \right)\Big|_{\textrm{top}}.
\end{equation}

\noindent Around the step $x$-location, the velocity-perturbation amplitude function displays typically two peaks along $y$; see Tufts \textit{et al.} \cite{Tufts17} and figure $12$ in Casacuberta \textit{et al.} \cite{Casacuberta22JFM}. On the one hand, the lower peak \cite{Tufts17} is linked to the locally formed near-wall streaks induced at the step \cite{Casacuberta21,Casacuberta22JFM}. On the other hand, the upper peak --that pre-exists upstream of the step-- is associated to the original crossflow instability. In order to fully isolate incoming perturbation effects, we measure the amplitude for all $x$ at the wall-normal location of the upper peak of the velocity-perturbation amplitude function, here referred to as \textit{top} (Eq.~\ref{eq:energy}). 

Thin solid lines in Fig.~\ref{fig:evol3D}(\textit{b}) characterize the evolution of $E^{\textrm{top}}_{\beta_{0}}$ in $x$ for the $h / \delta_{0} = 0.97$ step case (red line) and for the flat-plate reference case (black line). Their trends confirm that the stationary crossflow perturbation is stabilized locally downstream of the step. Dotted orange ($h / \delta_{0} = 0.59$) and dash-dotted blue ($h / \delta_{0} = 0.76$) lines in Fig.~\ref{fig:evol3D}(\textit{b}) additionally characterize the perturbation-energy evolution in steps smaller than the stabilising case ($h / \delta_{0} = 0.97$) characterized by thin solid red. The stabilizing trend does not manifest for these cases with smaller steps.

\subsection{Decomposition of the production term}

\begin{figure}
\begin{center}
\includegraphics{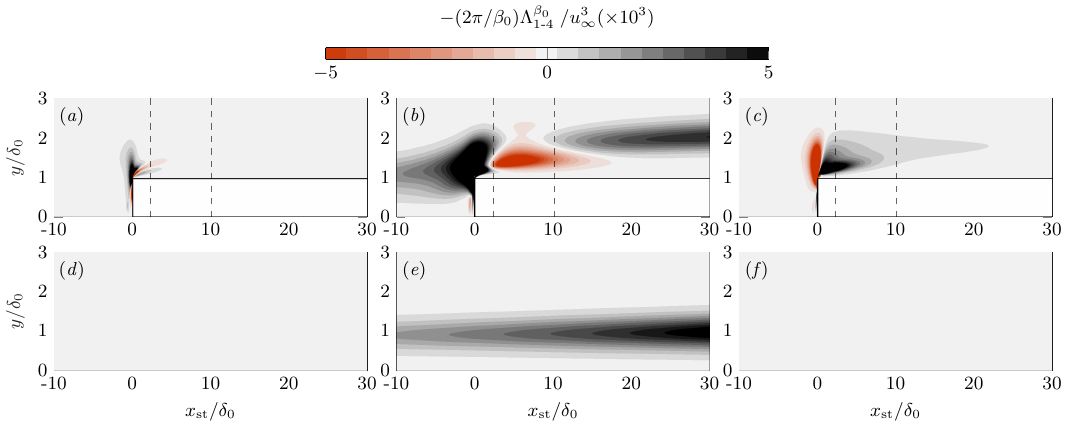}
\caption{\label{fig:I1toI4} {Spatial organization of the integrand of $I^ {\beta_{0}}_{1} + I^ {\beta_{0}}_{3}$(\textit{a},\textit{d}), $I^ {\beta_{0}}_{2}$(\textit{b},\textit{e}), $I^ {\beta_{0}}_{4}$(\textit{c},\textit{f}) in the step (top) and reference no-step (bottom) cases. Dashed black lines indicate $x$-range of \textit{top} perturbation decay shown in Fig.~\ref{fig:evol3D} (grey area).}}
\end{center}
\end{figure}

To elucidate the mechanism responsible for the CFI stabilization illustrated in Fig.~\ref{fig:evol3D}, the exchange of kinetic energy between the base flow and perturbations at the step is analyzed by means of the production term of the Reynolds-Orr equation, $P_{\beta_{0}} = -\int_{V} \hat{\bm{\upsilon}}^{\prime} \bm{\cdot} \left( \hat{\bm{\upsilon}}^{\prime} \bm{\cdot} \nabla \right) \bm{\upsilon}_{\textrm{B}}\;\textrm{d}V$ ($\S$ \ref{sec:decomP}). Here, the integration volume $V$ is chosen to encompass the step in $x$ and to extend towards the transverse boundaries in $z$ and the free-stream in $y$. Figure \ref{fig:I1toI4} portrays the integrands of $I^{\beta_{0}}_{m}, m = 1$-$4$, stemming from the decomposition of $P_{\beta_{0}}$ (Eq.~\ref{eq:firstPDecomp}), in the step (\textit{a}-\textit{c}) and reference no-step (\textit{d}-\textit{f}) cases. As expected, the term $I^{\beta_{0}}_{2}$ holds the dominant contribution to energy production in the no-step case, see Fig.~\ref{fig:I1toI4}(\textit{e}). This highlights the role played by the weak cross-stream pattern ($\bm{\upsilon}^{\prime}_{n}$) produced by the instability which, by displacing base-flow momentum, it enhances regions of streamwise-velocity deficit and excess ($\bm{\upsilon}^{\prime}_{t}$); consequently, the crossflow perturbation is amplified spatially \cite{Saric03}. In the presence of the step, the mechanism $I^{\beta_{0}}_{2}$ remains a dominant contribution in absolute value (Fig.~\ref{fig:I1toI4}(\textit{b})), albeit an enhancement of the mechanism associated to $I^{\beta_{0}}_{4}$ is captured \textit{locally} near the step corner (Fig.~\ref{fig:I1toI4}(\textit{c})). This latter feature has been reported as well in studies of near-wall streaks in unswept forward-facing-step flows \citep{Lanzerstorfer12}. 

Figure \ref{fig:I1toI4}(\textit{b}) shows that the dominant production term $I^{\beta_{0}}_{2}$ reverts in sign shortly downstream of the step, approximately from $x_{\textrm{st}} / \delta_{0} = 2.3$; see red contour in Fig.~\ref{fig:I1toI4}(\textit{b}). Following the key outcomes from the model problems, this essentially corresponds to the lift-up effect acting in a stabilizing manner, that is, by transferring kinetic energy from the perturbation field to the underlying flow. We refer to this mechanism as \textit{reverse} lift-up effect since, originally, the \textit{classic} lift-up effect was conceived as a mechanism responsible for actually destabilizing streamwise streaks through the action of cross-stream perturbations \citep{Ellingsen75,Landahl75,Landahl80}. The $x$-position where $I^{\beta_{0}}_{2}$ first reverts in sign approximately matches the location at which the crossflow perturbation decays in $x$ (i.e.\ $\textrm{d} E^{\textrm{top}}_{\beta_{0}} / \textrm{d}x < 0$) downstream of the step (Fig.~\ref{fig:evol3D}(\textit{b})). 

Downstream of the stabilizing region, the perturbation structures gradually re-organize  towards reference no-step conditions (see $\S$ \ref{sec:inOutPhase}) and a destabilizing influence of $I^{\beta_{0}}_{2}$ progressively sets in again (black contour in figure \ref{fig:I1toI4}(\textit{b})). In this flow environment, the \textit{strength} of $I^{\beta_{0}}_{2} > 0$ (Fig.~ \ref{fig:I1toI4}(\textit{b})) is lower than in reference conditions (Fig.~\ref{fig:I1toI4}(\textit{e})), implying that the transfer rate of kinetic energy towards the perturbation field is below reference no-step conditions. This reconciles with the reduced growth rate in $x$ of the crossflow perturbation after passing the region of $I^{\beta_{0}}_{2} < 0$ in the step case, evident in Fig.~\ref{fig:evol3D}(\textit{b}). When moving further downstream, far from the flow distortion introduced by the step, the growth rate of the crossflow perturbation eventually increases significantly and becomes closer to the no-step case.

\subsection{On the stabilizing or destabilizing contribution of the lift-up effect}\label{sec:inOutPhase}

\begin{figure}
\begin{center}
\includegraphics{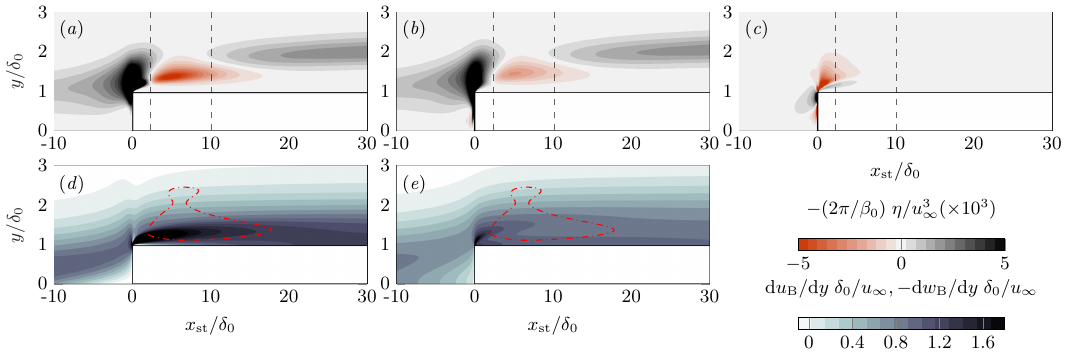}
\caption{\label{fig:I2terms} {Spatial organization of terms of Eq.~(\ref{eq:liftUp01}): $\eta = \kappa^{\beta_{0}}_{2}$ (\textit{a}), $\delta^{\beta_{0}}_{2}$ (\textit{b}), remainder (\textit{c}). Dashed lines indicate $x$-range of perturbation decay shown in Fig.~\ref{fig:evol3D} (grey area). Evolution of base-flow shear $\textrm{d}u_{\textrm{B}} / \textrm{d}y$(\textit{d}) and $-\textrm{d}w_{\textrm{B}} / \textrm{d}y$(\textit{e}) with isoline of $\Lambda^{\beta_{0}}_{2} \approx 0$ (dash-dotted red).}}
\end{center}
\end{figure} 

The fundamental construction of the four individual production terms (Eq.~\ref{eq:firstPDecomp}) provides a powerful means for gaining insights into mechanisms of growth or decay in these flows. To identify the origin of the aforementioned reversed action of the dominant lift-up effect ($I^{\beta_{0}}_{2}$) at the step, the following expressions are considered:  
\begin{equation}\label{eq:kappa}
\kappa^{\beta_{0}}_{2} = 2 \frac{\partial u_{\textrm{B}}}{\partial y} |\tilde{\upsilon}^{1}_{t}| |\tilde{\upsilon}^{2}_{n}| \cos \left( \varphi_{t} - \varphi^{\upsilon^{2}}_{n} \right),
\end{equation}
\begin{equation}\label{eq:delta}
\delta^{\beta_{0}}_{2} = -2 \frac{\partial w_{\textrm{B}}}{\partial y} |\tilde{\upsilon}^{3}_{t}| |\tilde{\upsilon}^{2}_{n}| \cos \left( \varphi_{t} - \varphi^{\upsilon^{2}}_{n} \right),
\end{equation}

\noindent namely the terms of Eq.~(\ref{eq:liftUp01}) that have the largest contribution to $I^{\beta_{0}}_{2}$ in the step case, as quantitatively demonstrated in Fig.~\ref{fig:I2terms}(\textit{a}-\textit{c}). From Eq.~(\ref{eq:kappa}), $\kappa^{\beta_{0}}_{2}$ is conceived as the contribution to lift-up ($I^{\beta_{0}}_{2}$) by which the wall-normal shear of the base-flow $u_{\textrm{B}}$ (with $\partial u_{\textrm{B}} / \partial y > 0$) amplifies $\bm{\upsilon}_{t}$ in the direction $x$. Similarly, $\delta^{\beta_{0}}_{2}$ expresses the contribution by which the wall-normal shear of the base-flow $w_{\textrm{B}}$ (with $\partial w_{\textrm{B}} / \partial y < 0$) amplifies $\bm{\upsilon}_{t}$ in the direction $z$. 

The base-flow gradients $\textrm{d}u_{\textrm{B}} / \textrm{d}y$ and $\textrm{d}w_{\textrm{B}} / \textrm{d}y$ do not change sign in the flow regime dominated by the \textit{reverse lift-up effect}. This is shown in Fig.~\ref{fig:I2terms}(\textit{d},\textit{e}). A small region of flow reversal (i.e.\ $\textrm{d}u_{\textrm{B}} / \textrm{d}y < 0$) downstream at the step is localized at the step apex and no significant impact of this flow structure on the presently discussed mechanism can be identified. Therefore, the sign of both $\kappa^{\beta_{0}}_{2}$ and $\delta^{\beta_{0}}_{2}$, i.e.\ whether they are stabilizing or destabilizing contributions to $I^{\beta_{0}}_{2}$, is dictated by a unique and common factor, namely $\cos ( \varphi_{t} - \varphi^{\upsilon^{2}}_{n} )$. The latter evaluates the relative phase between the component of the cross-stream velocity perturbation in $y$, \textit{acting} on the wall-normal shears of the base flow, i.e.\ $\varphi^{\upsilon^{2}}_{n}$ (Eq.~\ref{eq:phaseNorm}), and the streamwise-velocity perturbation component, i.e.\ $\varphi_{t}$ (Eq.~\ref{eq:phaseTang}). In short, the stabilizing or destabilizing contribution of the lift-up effect at the step is established by the relative arrangement of cross-stream- and streamwise-velocity perturbations.

\begin{figure}
\begin{center}
\includegraphics{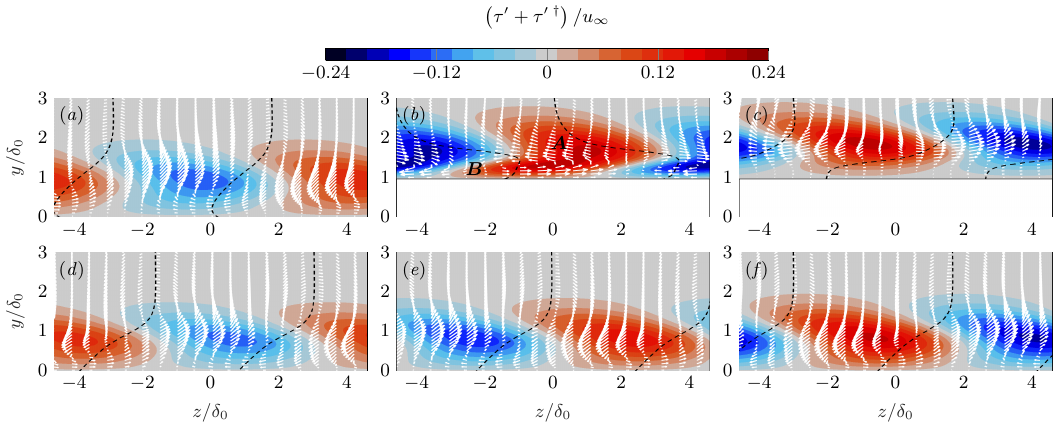}
\caption{\label{fig:physInterp} {Streamwise-velocity perturbation (color map) in $y$-$z$ planes for the step (top) and reference no-step (bottom) cases at $x_{\textrm{st}} / \delta_{0} = -10$(\textit{a},\textit{d}), $5$(\textit{b},\textit{e}), $20$(\textit{c},\textit{f}). In-plane organization of the cross-stream-velocity perturbation ($\bm{\upsilon}^{\prime}_{n}$) depicted as white arrows. Dashed black segregates regions of perturbation upwash and downwash (i.e.\ $\upsilon^{\prime 2}_{n} = \bm{\upsilon}^{\prime}_{n}(2) = 0$).}}
\end{center}
\end{figure}

To provide a conceptual model of CFI stabilization by the step, the organization of the fields $\bm{\upsilon}^{\prime}_{t}$ and $\bm{\upsilon}^{\prime}_{n}$ is examined in relation to the identified dominant factor, $\cos ( \varphi_{t} - \varphi^{\upsilon^{2}}_{n} )$, in Eqs.~(\ref{eq:kappa}) and (\ref{eq:delta}). Figure \ref{fig:physInterp} shows streamwise-velocity perturbation $\tau^{\prime}$ (Eq.~\ref{eq:tanDef}) represented by color contour with white arrows illustrating the organization of the counter-rotating cross-stream perturbation ($\bm{\upsilon}^{\prime}_{n}$) in $y$-$z$ planes for the step case and the reference no-step case. Upstream of the step (Fig.~\ref{fig:physInterp}(\textit{a})), the perturbation behaviour qualitatively resembles reference no-step conditions (Fig.~\ref{fig:physInterp}(\textit{d}-\textit{f})); that is, the action of perturbation upwash (i.e.\ $\upsilon^{\prime 2}_{n} = \bm{\upsilon}^{\prime}_{n}(2) > 0$) dominates in regions of streamwise-velocity deficit (i.e.\ $\tau^{\prime} < 0$) and vice-versa. This interplay between perturbation components highlights the essence of the classic lift-up effect, namely the cross-stream velocity perturbations redistribute base-flow momentum by displacing low-momentum fluid upward and high-momentum fluid downward. Therefore, regions of streamwise-momentum deficit and excess are enhanced spatially. In such scenario, the cross-stream- and streamwise-velocity perturbation structures act \textit{out-of-phase}, i.e.\ $|\varphi_{t} - \varphi^{v^{2}}_{n}| > \pi / 2$, resulting in perturbation growth (Fig.~\ref{fig:evol3D}(\textit{b})). 

In the close vicinity of the step (Fig.~\ref{fig:physInterp}(\textit{b})), the perturbation organization is altered significantly, as compared to reference conditions at the same $x$-location (Fig.~\ref{fig:physInterp}(\textit{e})). At first glance, vigorous perturbation amplification in $x$ is captured near the wall (region labelled as ``B''). This is ascribed to the inception of newly-formed streaks close to the wall, as  described above in $\S$ \ref{sec:overviewStep}. However, the scope of this work is the behaviour of the original CFI that develops further from the wall (region labelled as ``A''). In region ``A'' in Fig.~\ref{fig:physInterp}(\textit{b}), perturbation upwash (i.e.\ $\upsilon^{\prime 2}_{n} = \bm{\upsilon}^{\prime}_{n}(2) > 0$) dominates in regions of streamwise-velocity excess (i.e.\ $\tau^{\prime} > 0$) and vice-versa. Thus, the action of the cross-stream velocity perturbation ($\bm{\upsilon}^{\prime}_{n}$) now weakens the incoming regions of streamwise-momentum deficit and excess and hence reduces the amplitude of $\bm{\upsilon}^{\prime}_{t}$. Following the discussion provided in $\S$ \ref{sec:inOutPhase}, the cross-stream- and streamwise-velocity perturbation structures now act \textit{in-phase}, i.e.\ $|\varphi_{t} - \varphi^{v^{2}}_{n}| < \pi / 2$, and $I^{\beta_{0}}_{2} < 0$. Therefore, the process is locally stabilizing and a decay of the perturbation energy in $x$ is consequently monitored (Fig.~\ref{fig:evol3D}(\textit{b})). Eventually when moving further downstream of the step, the cross-stream- and streamwise-velocity perturbation structures re-organize towards undisturbed (i.e.\ no-step) conditions and they act \textit{out-of-phase} again (Fig.~\ref{fig:physInterp}(\textit{c})). 

\begin{figure}
\begin{center}
\includegraphics{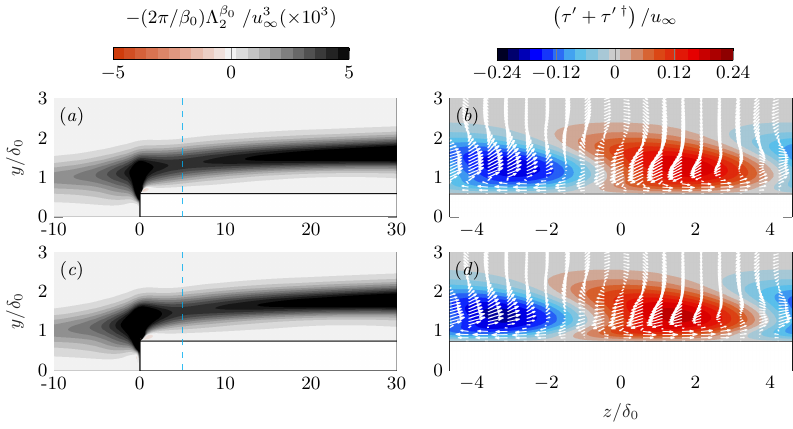}
\caption{\label{fig:physInterpExtraStep} {Spatial organization of the integrand of $I^{\beta_{0}}_{2}$ in additional steps of height $h / \delta_{0} = 0.59$(\textit{a}) and $0.76$(\textit{c}). Corresponding behaviour of the streamwise-velocity perturbation (color map) and in-plane organization of the cross-stream-velocity perturbation (white arrows) at $x / \delta_{0} = 5$ (indicated by dashed cyan line) for $h / \delta_{0} = 0.59$(\textit{b}) and $0.76$(\textit{d}).}}
\end{center}
\end{figure}

The new near-wall perturbation rolls induced and enhanced at the step (white arrows in Fig.~\ref{fig:physInterp}(\textit{b})) take over the incoming cross-stream perturbation motion and induce locally a reverse lift-up effect by acting \textit{against} (i.e.\ dampening) the incoming crossflow perturbation. Based on the current analysis, these near-wall rolls accompanying the near-wall streaks induced at the step appear to be the main step-flow feature responsible for the stabilization imparted by the step in the DNS of Casacuberta \textit{et al.} \cite{Casacuberta22JFM} and potentially in the experiments of Rius-Vidales and Kotsonis \cite{Rius21}.

The model of perturbation interaction described above is applicable to steps of different height and shape. At present, this is exemplified by considering additional steps with $h / \delta_{0} = 0.59$ and $0.76$. Both these step geometries were reported previously to act destabilizing \cite{Casacuberta22JFM} and thus to increase the amplitude of incoming CFI upon interaction. Conformably, here it is reported that $I^{\beta_{0}}_{2} > 0$ for all $x$ around these additional steps, see Fig.~\ref{fig:physInterpExtraStep}(\textit{a},\textit{c}). Moreover, in line with the discussion above, the qualitative interplay between perturbation components features no-step conditions, i.e.\ $\bm{\upsilon}_{t}$ and $\bm{\upsilon}_{n}$ act \textit{out-of-phase} (Fig.~\ref{fig:physInterpExtraStep}(\textit{b},\textit{d})). In the step geometries portrayed in Fig.~\ref{fig:physInterpExtraStep}, the near-wall rolls induced at the step are significantly weaker than in the main step case analyzed in this article (Fig.~\ref{fig:physInterp}(\textit{b})) and by their topology and organization they do not reverse the sense of energy production at the step.

While these new insights provide a basic understanding of CFI stabilization by a surface feature, similar mechanisms have been observed in other flow problems involving streaky perturbations. For instance, within the scope of non-modal growth in unswept convex surfaces, Karp and Hack \cite{Karp18} report a new pair of near-wall perturbation rolls which ``work against [pre-existing] streaks as they lift low-speed flow at the locations of the positive streaks and push high-speed flow towards the negative streaks.'' Sescu and Afsar \cite{Sescu18} investigate the stabilization of Görtler vortices by streamwise wall deformation, which is pointed out to be a control strategy even more efficient than an analogous BS arrangement in some cases. In words of Sescu and Afsar \cite{Sescu18}, the role of the surface deformations is to ``weaken the lift-up effect'' by correspondingly accelerating and decelerating fluid particles \cite{Sescu18}. The passive control mechanism of Sescu and Afsar \cite{Sescu18} appears to have similarities to the observations in the current work. 

\section{Conclusions}\label{sec:conclusions}

Recently, Rius-Vidales and Kotsonis \cite{Rius21} observed experimentally that spanwise-invariant surface (forward-facing) steps have the potential to delay crossflow-induced laminar-turbulent transition on swept-wing flow. In parallel, by means of DNS, Casacuberta \textit{et al.} \cite{Casacuberta22JFM} identified significant stabilization of stationary crossflow vortices by a particular design of a forward-facing step. The underlying mechanism responsible for the passive stabilization reported by Casacuberta \textit{et al.} \cite{Casacuberta22JFM}, here defined as \textit{reverse lift-up effect}, has been identified and scrutinized in this present article.

The lift-up effect is a well-known destabilizing flow mechanism \cite{Moffatt67,Ellingsen75,Landahl75,Landahl80} which drives the amplification of perturbations in a broad range of shear-flow scenarios. We have elaborated upon the existence of a mechanism analogous to the lift-up effect, but acting stabilizing and \textit{reversely} to its classic conception. This may be understood as follows: the \textit{classic} lift-up effect \cite{Moffatt67,Ellingsen75,Landahl75,Landahl80} entails a three-dimensional cross-stream perturbation superimposed on a shear layer which, by redistributing low- and high-momentum fluid, induces inherently streamwise perturbation streaks (i.e.\ flow-aligned regions of momentum deficit and excess). It must be noted that the appearance of the lift-up effect only requires a cross-stream perturbation in an otherwise unperturbed base flow. However, under the condition that a flow instability pre-exists and carries stream-tangent perturbations, the momentum redistribution by the cross-stream perturbation component may be altered locally (eg. via rapid variation of surface geometry or base flow) such that it acts by quenching the pre-existing regions of streamwise-momentum deficit and excess. In short, when such a \textit{reverse lift-up effect} is active, high-momentum fluid is transported towards an incoming low-speed streak and low-momentum fluid is transported towards an incoming high-speed streak. Thus, the streak system as a whole is attenuated and the flow exhibits a tendency towards returning to its original (unperturbed) laminar state. Such a (linear inviscid) algebraic-growth type of effect has the potential to induce a significant decay of perturbation kinetic energy locally in space or time. Whether a classic or a reverse lift-up effect dominates, is dictated essentially by the phase difference between cross-stream- and streamwise-velocity perturbations, inducing either a constructive or a destructive wave-like interference in terms of kinetic-energy exchange.   

We have developed a theoretical framework to characterize corresponding energy-transfer mechanisms between the laminar base flow and perturbation fields; the framework is applicable to generic three-dimensional flows with one invariant spatial direction. To introduce and exemplify the framework and the novel concept of a reverse lift-up effect, two simple model problems have been discussed first. Namely, \textit{optimal} perturbations in plane Poiseuille flow and steady blowing-suction in two-dimensional boundary-layer flow. While they are canonical examples of perturbation amplification through the classic lift-up effect, it is shown that the perturbation equations support additionally a stabilizing (reverse) lift-up effect in these model problems. In this line, the emergence of a reverse lift-up effect follows naturally from the original model of the lift-up effect presented by Ellingsen and Palm \cite{Ellingsen75} if an \textit{initial} perturbation streak field is assumed. Following the analysis of the model problems, it is found that the lift-up effect dominates the mechanisms of perturbation interaction between a pre-existing stationary crossflow instability and a forward-facing step; whether the classic or a reverse lift-up effect dominates depends upon the choice of step height. The reverse lift-up effect is always localized. Nonetheless, the subsequent slow spatial relaxation of perturbations towards no-step conditions yields a large area of reduced growth rate of the crossflow instability.

The identification of a passive geometry-induced mechanism leading to (primary wavelength) stationary-crossflow perturbation stabilization is a promising finding for flow control research and aircraft design. 
A significant reduction of the amplitude of stationary crossflow vortices, which drive the process of laminar-turbulent transition in swept-wing flow, may be achieved by appropriate design of surface features.  Moreover, the mechanism characterized in this work does not require previous knowledge of the wavelength or the perturbation phase upstream of the surface feature. Therefore, it may be applied successively throughout an aerodynamic surface for an overall enhancement of its underlying benefit without need for active phase calibration. By the time of completion of this work, the authors obtained numerical \cite{Westerbeek23} and experimental evidence of stabilization of stationary crossflow vortices by a smooth (as opposed to sharp) surface feature, leading to significant transition delay as well.

\acknowledgements

This work was carried out on the Dutch national e-infrastructure with the support of SURF
Cooperative. The authors would like to acknowledge the support of the European Research Council through
StG no. 803082 GLOWING. Moreover, the authors would like to express their gratitude to Dr. Valerio Lupi for his assistance on the stability code for channel flows, as well as to Dr. A.F. Rius-Vidales, Dr. Giulia Zoppini, Sven Westerbeek, Marina Barahona, and Srijit Sen for fruitful discussions on the topic. Debating with a curious and passionate student is, and will always be, a main source of scientific innovation.

\section{Appendix A: A historical perspective on the classic and the reverse lift-up effect}\label{ref:appendixA}

Ellingsen and Palm \cite{Ellingsen75} and Landahl \cite{Landahl75,Landahl80} are credited mainly for formalizing the lift-up effect, the destabilizing flow mechanism responsible for the widespread presence of streaky structures in many shear-flow configurations \cite{Brandt14}. At first glance, the novel concept of a stabilizing (reverse) lift-up effect introduced in this present article seemingly opposes their main conclusion. This is not the case; in Appendix A, we elaborate upon the fact that the reverse lift-up effect may be regarded as an additional solution to the model of Ellingsen and Palm \cite{Ellingsen75}. This is expanded upon in the following manner: Ellingsen and Palm \cite{Ellingsen75} assume a parallel (base) flow with $\bm{\upsilon}_{\textrm{B}} = [u_{\textrm{B}}(y) \; 0 \; 0]^{\textrm{T}}$, which is incompressible, not stratified, and confined between two parallel walls. They consider the first component of the linearized inviscid perturbation equation; i.e.
\begin{equation}\label{eq:EP1}
\frac{\partial \hat{u}^{\prime}}{\partial t} + \hat{v}^{\prime} \frac{\partial u_{\textrm{B}}}{\partial y} = 0.
\end{equation}

\noindent For an $x$-invariant \cite{Ellingsen75} cross-stream perturbation, $\hat{v}^{\prime}$, the solution to Eq.~(\ref{eq:EP1}) between a time $t = t_{0}$ and $t_{1}$ is
\begin{equation}\label{eq:EP2}
\hat{u}^{\prime} = \hat{u}^{\prime}_{0} - \hat{v}^{\prime} \frac{\partial u_{\textrm{B}}}{\partial y} \Delta t,
\end{equation}

\noindent where $\hat{u}^{\prime}$ expresses a streamwise-velocity perturbation, $\Delta t = t_{1} - t_{0}$, $\hat{u}^{\prime}_{0} = \hat{u}^{\prime}(t = t_{0})$, and $\hat{v}^{\prime} \neq \hat{v}^{\prime}(t)$ under the present formulation \cite{Ellingsen75}. Ellingsen and Palm \cite{Ellingsen75} state that Eq.~(\ref{eq:EP2}) ``shows that $\hat{u}^{\prime}$ increases linearly with time.'' While the observation of Ellingsen and Palm \cite{Ellingsen75} is that in this context $\hat{u}^{\prime}$ evolves algebraically (as opposed to exponentially) in time, in fact $\hat{u}^{\prime}$ and hence the kinetic perturbation energy, $E_{V}$, might as well decay (algebraically) in time. 

This is exemplified as follows. Consider a wave-like perturbation ansatz; i.e.\ \mbox{$\hat{u}^{\prime} = \tilde{u} \textrm{e}^{\textrm{i} \beta_{0} z} + \textrm{c.c.}$} and \mbox{$\hat{v}^{\prime} = \tilde{v} \textrm{e}^{\textrm{i} \beta_{0} z} + \textrm{c.c}$}, where $\tilde{u} = |\tilde{u}| \textrm{e}^{\textrm{i} \varphi^{u}}, \tilde{v} = |\tilde{v}| \textrm{e}^{\textrm{i} \varphi^{v}}$, $z$ denotes the spanwise direction, $\beta_{0}$ indicates the perturbation wavenumber in the direction $z$, and it is assumed that $\varphi^{u}(t) = \varphi^{u}(t = t_{0})$ for $t_{0} \leq t \leq t_{1}$. Upon introducing these perturbation expressions into Eq.~(\ref{eq:EP2}), two conditions are retrieved. Namely, $|\tilde{u}| = |\tilde{u}|_{0} - \cos \left(\varphi^{v} - \varphi^{u} \right) |\tilde{v}| \Delta t \partial u_{\textrm{B}} / \partial y$ and $\varphi^{v} - \varphi^{u} = 0, \pi$. Therefore, the original model of Ellingsen and Palm \cite{Ellingsen75} admits two main solutions:
\begin{align} \label{eq:1975classic}\tag{3.2,\textit{a}}
& |\tilde{u}| = |\tilde{u}|_{0} + |\tilde{v}| \frac{\partial u_{\textrm{B}}}{\partial y} \Delta t
& \textrm{(\textit{classic} lift-up effect)}, 
\\ \nonumber 
& |\tilde{u}| = |\tilde{u}|_{0} - |\tilde{v}| \frac{\partial u_{\textrm{B}}}{\partial y} \Delta t 
& \textrm{(\textit{reverse} lift-up effect)}. \label{eq:1975reverse}\tag{3.2,\textit{b}}
\end{align}

\noindent Ellingsen and Palm \cite{Ellingsen75} write that ``we therefore deduce [from Eq.~\ref{eq:EP2}] that the base flow $u_{\textrm{B}}(y)$ is unstable to this kind of infinitesimal disturbance [i.e.\ a prescribed cross-stream perturbation].'' However, Eqs.~(\ref{eq:1975classic}) and (\ref{eq:1975reverse}) highlight that, locally in time, the flow field may be actually unstable (i.e.\ $\textrm{d} E_{V} / \textrm{d}t > 0$) by a \textit{classic} lift-up effect or stable (i.e.\ $\textrm{d} E_{V} / \textrm{d}t < 0$) by a \textit{reverse} lift-up effect to a cross-stream-velocity perturbation ($\hat{v}^{\prime}$) if a streamwise-velocity perturbation ($\hat{u}^{\prime}$) pre-exists. 

\bibliography{apssamp}

\end{document}